\documentclass[a4paper,11pt]{article}
\pdfoutput=1 

\usepackage{jcappub} 

\def \RXJ {RXJ1131-1231~}
\def \LCDM {\ensuremath{\Lambda\text{CDM}}~}
\def \mkeV {2}
\def \mth {\ensuremath{m_{\text{TH}}}~}

\title{Lensing substructure quantification in \RXJ: A \mkeV keV lower bound on dark matter thermal relic mass}

\author[a,b]{Simon Birrer,}
\emailAdd{simon.birrer@phys.ethz.ch}
\author[a]{Adam Amara,}
\emailAdd{adam.amara@phys.ethz.ch}
\author[a]{Alexandre Refregier}
\emailAdd{alexandre.refregier@phys.ethz.ch}

\notoc

\affiliation[a]{Institute for Astronomy, Department of Physics, ETH Zurich\\ Wolfgang-Pauli-Strasse 27, 8093, Zurich, Switzerland}
\affiliation[b]{Department of Physics and Astronomy, University of California, Los Angeles\\ 475 Portola Plaza, Los Angeles, CA 90095-1547, USA (present address)}

\abstract{
We study the substructure content of the strong gravitational lens \RXJ through a forward modelling approach that relies on generating an extensive suite of realistic simulations.
We use a semi-analytic merger tree prescription that allows us to stochastically generate substructure populations whose properties depend on the dark matter particle mass. These synthetic halos are then used as lenses to produce realistic mock images that have the same features, e.g. luminous arcs, quasar positions, instrumental noise and PSF, as the data. We then analyse the data and the simulations in the same way with summary statistics that are sensitive to the signal being targeted and are able to constrain models of dark matter statistically using Approximate Bayesian Computing (ABC) techniques. In this work, we focus on the thermal relic mass estimate and fix the semi-analytic descriptions of the substructure evolution based on recent literature. We are able, based on the HST data for \RXJ, to rule out a warm dark matter thermal relic mass below \mkeV keV at the 2$\sigma$ confidence level.
}

\arxivnumber{1702.00009 }

\begin{document}
\maketitle
\flushbottom

\section{Introduction}

Dark matter is one of the core ingredients in cosmology and evidence for its existence can be found in many areas including the cosmic microwave background (CMB) temperature anisotropies \citep[][]{PlanckCollaboration:2015p9875}, galaxy clustering \citep{Anderson:2014p13626}, weak gravitational lensing \citep{Heymans:7p13831, TheDarkEnergySurveyCollaboration:2015p13765}, supernovae of type Ia \citep[e.g.][]{Riess:1998p14437,Perlmutter:1999p14481} and clusters \cite{Haan:2016p15036}. The smallest scale phenomenological tests come from the Lyman-alpha forest \citep[see e.g.,][]{Viel:2005p13838, Seljak:2006p9712}. The standard paradigm is that dark matter consists of cold particles that interact only through gravity.
On large cosmological scales this model makes predictions that remain consistent with data. However on small, sub-galactic, scales tensions between theoretical expectation and observational measurements seem to appear.
Discrepancies have been reported predominantly in the number, phase space densities and density profiles when comparing simulations of dark matter substructure with observations of luminous satellite galaxies in our Milky Way (MW) see e.g. \citep[][]{Kauffmann:1993p5842, Klypin:1999p4649, Moore:1999p4657, Kravtsov:2010p14006, BoylanKolchin:2011p4244} and a recent review \cite{DelPopolo:2016p14495}. There also remain open questions about the inner slopes in galaxy clusters \citep[e.g.][]{Sand:2003p15039}. These tensions attract attention due to the fact that some of these features could be explained by introducing dark matter particle properties and interactions beyond those of cold dark matter (CDM) of the standard cosmological model.

Warm dark matter (WDM), a thermal relic with a free streaming length, has been proposed to solve the discrepancies in the abundance of small scale structure \citep[][]{Bode:2001p13951, Abazajian:2006p13972}. 
The empirical constraints on this model, from the Lyman-$\alpha$ forest \citep[][]{Viel:2013p13930}, lead to a lower bound on the thermal relic mass $\mth = 3.3$keV at the 2$\sigma$ confidence level. The latest constraints from \cite{Irsic:2017p15429} results to $\mth = 5.3$keV based on their chosen priors on the thermal evolution. Allowing for a non-smooth evolution of the temperature of the IGM with sudden temperature changes reduces their lower limit to $\mth = 3.5$keV. The limits obtained from dwarf galaxy counts also disfavors particle masses below $\mth = 2.3$keV at the 2$\sigma$ confidence level \citep[][]{Polisensky:2011p14007, Kennedy:2014p14040}. Another important probe of substructure properties is strong gravitational lensing \citep[][]{Metcalf:2001p9744, Keeton:2009p9737, Moustakas:2009p9694}. For example, strong lensing has been used to detect luminous and dark substructure in strong lens systems \citep[][]{Koopmans:2005p8841, Vegetti:2009p9255}. This has led to the detection of individual clumps down to masses of roughly $2 \times 10^8 M_{\odot}$ \citep[][]{Vegetti:2010p9515, Vegetti:2012p4937, Hezaveh:2016p13593}. In these studies the approach is to detect substructure on an object by object basis. However, to make precision measurements of WDM matter properties, we need to move to statistical methods that are able to capture information from the collective action of the population of subhalos within a parent halo. Statistical approaches to quantify substructure have been done in simple lens configurations, for instance by analyzing flux ratios in multiple imaged lensed quasars \citep[see e.g.,][]{Metcalf:2002p9547, Amara:2006p4715, Metcalf:2012p5467, Xu:2015p9516, Nierenberg:2017p15476}. Anomalous flux ratios have been reported relative to a prediction based on simple smooth lens models, known as flux-ratio anomalies. Constraints on warm dark matter from weak lensing in anomalous quadruple lenses \citep[][]{Inoue:2015p15467} lead to $\mth = 1.3$keV under the assumption that the lens profile is perfectly known. These approaches requires multiple strong lens systems as the statistics is rather weak for a single lens system as well as strong assumptions about the lens mass profile.

In this paper, we take a different statistical approach to quantifying substructure in strong gravitational lenses. Our method is based on an extensive forward modelling scheme that relies on simulating lensing imaging data using a suite of halos, where the statistics of the subhalo population are set by the dark matter properties.
These generated mocks are tuned to be similar to the data \citep[see e.g.][for other application of forward modelling in cosmology]{Kacprzak:2012p14788, Mesinger:2011p14795, Reinecke:2006p14798, Peterson:2015p14808, Juin:2007p14830, Pires:2009p14840, Dietrich:2010p14848, Marian:2012p14852, Heymans:2006p14883, Massey:2007p14886, Bridle:2009p14888, Refregier:2014p14915, Bruderer:2016p14919}. 
A simple feature of a forward modelling approach is that the same analysis tools are run on both simulations and the data. This then allows us to use an approach known as Approximate Bayesian Computing (ABC) \cite{Rubin:1984p15259, Diggle:1984p15276, Tavare:1997p15227, Turner:2012p14944, Liepe:2014p14925} to perform Bayesian inference calculation even for cases where the likelihood cannot be computed. This is because the variability in the mocks allows us to propagate errors to the uncertainties on the parameters being estimated. A key step to this approach is to find diagnostic measures, known as summary statistics, that are sensitive to the signal we wish to target, which is substructure in our case. A recent ABC framework for cosmological purpose has been implemented by \cite{Akeret:2015p15286}.
We apply our method to the quadrupole lens system \RXJ, that was discovered by \cite{Sluse:2003p8680}, to quantify the substructure content and to differentiate between different dark matter models.

This paper is structured as follows: In section \ref{sec:smooth_lens_model}, we describe the smooth lens model and source reconstruction of the lens \RXJ. In section \ref{sec:simulations}, we describe our substructure modeling and how we create realistic mock lenses for different dark matter models. We then describe how we perform a substructure analysis and how we compare the statistical features in section \ref{sec:model_comparison}. In section \ref{sec:dm_constraints}, we present our dark matter model constraints based on the analysis of the lens \RXJ based on hundreds of different simulations. We discuss the results in section \ref{sec:discussion_substr}, compare them to the literature and discuss possible extensions to this work. Further technical details about the analysis are provided in the appendices.
Throughout this work, we assume a flat \LCDM cosmological model with $H_0 = 69.31$ km~s$^{-1}$Mpc$^{-1}$, $\Omega_b = 0.049$, $\Omega_m=0.315$, $\sigma_{8}=0.829$ and $n_s = 0.968$.

\section{Smooth lens model of \RXJ}\label{sec:smooth_lens_model}
In this work, we focus on substructure inferences based on the strong lens system \RXJ. The redshift of the lens is $z_l = 0.295$ and of the background quasar source $z_s = 0.658$. These were determined spectroscopically by \cite{Sluse:2003p8680}. Throughout this work, we make use of the \texttt{MultiDrizzle} product from the HST archive in bands ACS WFC F814W and F555W. We use a 160$^2$ pixel image centered at the lens position with pixel scale 0.05". This corresponds to a FOV of 8"$\times$ 8".

In targeting the signatures of substructure we separate our analysis into global features and small scale structure, where we focus on the latter to measure dark matter properties. Furthermore, two components need to be set for the global model, the lens and the source. 
For the smooth lens model, we use the same model parameters as in \cite{Birrer:2016p13196}. This model is composed of an ellipsoidal power law mass distribution, an SIS profile centered on the visible lens substructure and external shear parameters. Additionally, we use shapelet potential perturbations, as introduced in \cite{Birrer:2015p11550} based on the shapelet basis set \cite{Refregier:2003p8153}. We set the scale parameter of the lens model shapelets to be the Einstein radius of the lens.
In total, we use 21 shapelet parameters (corresponding to $n_{\text{max}}=5$), which enables us to model the lens mass distribution down to 0.4" resolution.

For the global luminous source structure, we use shapelets up to order $n_{\text{max}}=50$ with scale parameter $\beta=0.18"$ centered at the position of the quasar, as described in \cite{Birrer:2015p11550}. In a second step, we increase the source resolution with adaptive shapelets. We do this by identifying multiple distinct features in the residual map that correspond to source structure on smaller scales. We traced back the residual features to the source plane and identify 15 distinct regions corresponding to flux coming from small scale structure. For each of these specific regions, we place additional shapelets with order $n_{\text{max, clump}}=3$ and scale $\beta_{\text{clump}}$ set to the magnified resolution limit of the data. The source reconstruction is performed independently for the two imaging bands, but the common configuration and order of the basis sets is used for the source.

We reconstruct the two HST ACS filter bands F814W and F555W simultaneously and combine the likelihood of the smooth lens model given the imaging data. We use the framework presented in \cite{Birrer:2015p11550} to fit the high dimensional non-linear parameter space using a Particle-Swarm Optimization (PSO) method \cite{Kennedy:2001p8447}. The best fit lens model result has a reduced $\chi^2$ value of $\chi^2_{\text{red}} = 1.03$. 

In modeling the global feature that do not come from substructure, care needs to be taken in two respects. The first is that the smooth lens model needs to be sufficiently flexible to capture the scales that are larger than the scales where dark matter substructure has an impact. In appendix \ref{app:sys_smooth_lens}, we show how the choice of the smooth model can affect the substructure analysis. The second consideration is that the substructure signal depends on the source surface brightness variation. This means that we need to ensure that the simulations have the right level of small scale variation in the source to give reliable statistics in the mocks. We also need to allow the model to have sufficient small scale features as present in the data so as to not lead to biases. In appendix \ref{app:sys_source_res}, we present illustrative examples and a discussion of how these model biases affect substructure signatures.

\section{Simulations}\label{sec:simulations}

We anchor our comparison of the statistical features in the image residuals with the features of different dark matter models on simulations. For our analysis, we target deflection angle perturbations arising from substructure within the lens relative to a smooth lens model. To investigate the statistical significance of a signal, we use multiple realizations of the same physical model (WDM, CDM) and different halo masses. We use semi-analytic descriptions to generate a large number of realizations of a given model. The details on how we produce the set of simulations and how we incorporate WDM in the semi-analytic description is given in this section.

\subsection{Dark matter substructure model} \label{sec:wdm_simulations}
To compute a sample of expected deflection perturbations, we use a model based on the extended Press-Schechter \cite{Press:1974p410, Bond:1991p341} formalism. In particular, we use a merger tree and a subhalo evolution and disruption prescriptions, tuned to N-body CDM simulations \citep[see e.g.][]{Lacey:1993p337, Parkinson:2008p31, Neistein:2008p15323, Tweed:2009p15335, Benson:2016p15322}.

\subsubsection{Power spectrum}\label{sec:power_spectrum}
We follow \cite{Viel:2005p13838} in describing a power spectrum for WDM. The relative transfer function $r_T(k)$, defined as the ratio of WDM to CDM model can be defined as
\begin{equation}
	r_T(k) = \left[P(k)_{\text{$\Lambda$WDM}}/P(k)_{\text{$\Lambda$CDM}} \right]^{1/2}.
\end{equation}
In the case of WDM, the relative transfer function is given by \cite{Bode:2001p13951}
\begin{equation}
	r_T(k) = \left[1 + \left(\alpha k \right)^{2\nu}\right]^{-5/\nu},
\end{equation}
where $\alpha$ is the scale of the free streaming break of the WDM particle and $\nu$ was fixed by \cite{Viel:2005p13838} to $\nu = 1.12$ and
\begin{equation}
	\alpha = 0.049 \left(\frac{m_{\chi}}{1\text{keV}}\right)^{-1.11} \left(\frac{\Omega_{\chi}}{0.25}\right)^{0.11} \left(\frac{h}{0.7}\right)^{1.22}h^{-1}\text{Mpc}.
\end{equation}
$m_{\chi}$ is the DM thermal relic mass and $\Omega_{\chi}$ is the normalized dark matter density. The mass variance $\sigma^2(M)$ can be computed as
\begin{equation}
	\sigma^2(M) = \frac{1}{2\pi^2} \int_0^{\infty} 4 \pi k^2 P(k) W^2(k|M)dk,
\end{equation}
which depends on the window function $W$ used. 

A sharp-$k$ filter can capture the impact of a truncated power spectrum (such as WDM) since the suppressed high $k$ modes leads to a flattening of $\sigma^2(M)$. We use the sharp-$k$ filter

\begin{equation}
	W(k|M) =
	\begin{cases}
    1 & \text{if } k \le k_s(M)\\
    0 & \text{if } k > k_s(M).
	\end{cases}
\end{equation}
A definition of the mass $M$ according to the filter scale $k_s$ and a clear physical relation on the $k_s(M)$ relation is missing. We follow \cite{Benson:2013p13572} and relate $k_s$ to the mass radius $R$ of a real space top-hat filter through a scale parameter $a$ with $k_s = a/R$ as
\begin{equation}\label{eq:M_R}
	M = \frac{4\pi}{3} \bar{\rho}R^3 = \frac{4\pi}{3} \bar{\rho} \left(\frac{a}{k_s}\right)^3.
\end{equation}
We set $a=2.5$ as stated by \cite{Benson:2013p13572} to predict the turnover in the halo mass function.

\subsubsection{Merger tree}
To generate stochastic merger trees based on the power spectrum and mass variance, we follow \cite{Benson:2013p13572}, which is based on the original merger tree for CDM structure of \cite{Parkinson:2008p31}. Using a sharp-$k$ filter to compute $\sigma^2(M)$ will change the halo mass function at high mass. To avoid over-predicting high mass halos, the barrier for collapse has to be increased by a factor of 1.197 \citep[see][]{Benson:2013p13572}.
Furthermore, WDM changes the collapse threshold below a characteristic mass \cite{Barkana:2001p14092}. Based on work of \cite{Barkana:2001p14092, Benson:2013p13572}, a threshold of collapse, $\delta_c(M,t)$, from WDM and CDM can be fit. 
For this, we adopt equation 7-10 of \cite{Barkana:2001p14092}. We do not, for simplicity, incorporate the non-Gaussian walks of a moving barrier in WDM, as proposed by \cite{Benson:2013p13572}. However, we set the resolution limit of the merger tree to the characteristic mass. For the subhalo population, we do not expect a significant impact of the simplified implementation. This description allows us to produce many realizations of halos and the in-fall histories of the accreted sub-halos.

\subsubsection{Mass-concentration relation}
The concentration of the dark matter halos depends on the mass assembly history. Halos that assemble earlier are more concentrated \citep[e.g.][]{Navarro:1997p8389, Wechsler:2002p14192, Ludlow:2013p2973}. Taking a mass-concentration-redshift relation based on CDM simulations may not reflect the fact that WDM halos form later and are therefore expected to have lower concentrations. We use the $c(M,z)$ relation of \cite{Duffy:2008p14292} for CDM cosmologies, namely 
\begin{equation}
	c = A_c\left(\frac{M}{M_{\text{pivot}}}\right)^{B_c}(1+z)^{C_c}.
\end{equation}
Throughout this work, we set $A_c = 5.22$, $B_c = -0.072$, $C_c = -0.42$ and $M_{\text{pivot}} = 2 \cdot 10^{12} M_{\odot}$, taken from \cite{Duffy:2008p14292}. In the WDM case, we set the $c(M,z)$ relation by a mapping of the same formation histories of CDM halos as described in \cite{Schneider:2015p14276}. Halos of different WDM models with the same formation history, but different masses, have the same concentration. This leads to a lower concentration in a WDM case compared to CDM as a result of the later formation of the WDM halo.

\subsubsection{Substructure evolution}
To describe the evolution of the subhalos from the time they are assigned to a more massive halo, the information in a merger tree is insufficient since subhalos get disrupted by tidal stripping and dynamical friction.

We use the description of \cite{Jiang:2016p14171}, which is based on \cite{vandenBosch:2005p14184}, to track the average mass-loss rate, $\dot{m}$, of a dark matter subhalo expected by tidal stripping. The parameterized form of the subhalo decay $\dot{m}$ depends on the instantaneous mass ratio $m/M$ of the subhalo $m$ and the parent halo $M$ and the dynamical time $\tau_{\text{dyn}}$ of the halo as
\begin{equation}
	\dot{m} = - A_{ts} \frac{m}{\tau_{\text{dyn}}}\left(\frac{m}{M}\right)^{\zeta},
\end{equation}
where $A_{ts}$ and $\zeta$ describe the normalization and mass dependents of the subhalo decay. Throughout this work, we use $\tau_{\text{dyn}}=0.1 H^{-1}$. We take the parameter values derived from \cite{Giocoli:2008p312}, $A_{ts}=1.54$ and $\zeta = 0.07$. We do not introduce scatter in this relation \citep[as done by][]{Jiang:2016p14171} as we expect that the scatter introduced by the merger tree will have this effect.

Dynamical friction has also a significant impact on the subhalo disruption. We use the description of \cite{BoylanKolchin:2008p28}, which provides a fitting formula for this process based on simulations. The number of dynamical time scales to fully merge the subhalo into the parent $n_{\text{dyn}}$ is given by \cite{BoylanKolchin:2008p28} as
\begin{equation} \label{eqn:dyn_friction}
	n_{\text{dyn}} \equiv \frac{\tau_{\text{df}}}{\tau_{\text{dyn}}} = A_{df} \frac{(M/m)^{b_{df}}}{\ln(1+M/m)}\exp[c_{df}\eta]
\end{equation}
where $\eta = j/j_c(E)$ is the orbital circularity. We ignore the dependence on orbital energy. The fitting parameters are tuned to simulations and given by \cite{BoylanKolchin:2008p28} as $A_{df}=0.216$, $b_{df}=1.3$ and $c_{df}=1.9$. \cite{Zentner:2005p290} showed that the circularity $\eta$ of subhalos at infall can be well fit by
\begin{equation} \label{eqn:orbital_circ}
	P(\eta) \propto \eta^{1.22}(1-\eta)^{1.22}.
\end{equation}
We draw from this distribution for every subhalo at in-fall to compute the dynamical friction time according to equation \ref{eqn:dyn_friction}. The scale factor of disruption due to dynamical friction $a_{\text{df}}$ is then given by
\begin{equation}
	a_{\text{df}} = a_{\text{infall}}\exp\left[n_{\text{dyn}}/10\right].
\end{equation}
Subhalos whose dynamical friction time $a_{\text{df}}$ is shorter than that needed to survive to the redshift of interest (in our case to the redshift of the lens $a_{\text{lens}}$) are assumed to be fully disrupted.

For the evolution of the subhalo internal structure, we use the description of \cite{Penarrubia:2008p14342, Penarrubia:2010p14353}. The ratio of $y=r_{\text{max}}/r_{\text{max,infall}}$ and $y=v_{\text{max}}/v_{\text{max,infall}}$ are both parameterized as a function of mass loss $x=m/m_{\text{infall}}$ as
\begin{equation}
	y(x) = \frac{2^{\alpha}x^{\beta}}{(1+x)^{\alpha}}
\end{equation}
where $\alpha$ and $\beta$ are fitting coefficients. The coefficients from \cite{Penarrubia:2008p14342, Penarrubia:2010p14353} for $y=r_{\text{max}}/r_{\text{max,infall}}$ are $\alpha=-0.3$, $\beta=0.4$ and for $y=v_{\text{max}}/v_{\text{max,infall}}$ they are $\alpha=0.4$, $\beta=0.3$.

The semi-analytic prescription used to compute the disruption of halos does not have a specific concentration dependence. In the case of WDM, where the concentration of halos is expected to be lower than for CDM, ignoring the concentration as a parameter in the disruption process might slightly under-predict the strength of the disruption in the WDM case. For all of our conclusions, the bounds on the thermal relic mass of a WDM particle are conservative with respect to the subhalo disruption effect. In appendix \ref{app:subhalo_stat}, we show the subhalo mass function and maximum circular velocity function for different WDM models predicted with the stated descriptions.

The spacial position of the subhalos is computed based on the orbit at infall. The orbit is computed based on the orbital circularity distribution (see equation \ref{eqn:orbital_circ}) and provides a radial distribution function from which we draw from. Our tests show that the radial distribution of substructure mass follows the NFW profile of the parent halo.

\begin{figure*}
  \centering
  \includegraphics[angle=0, width=145mm]{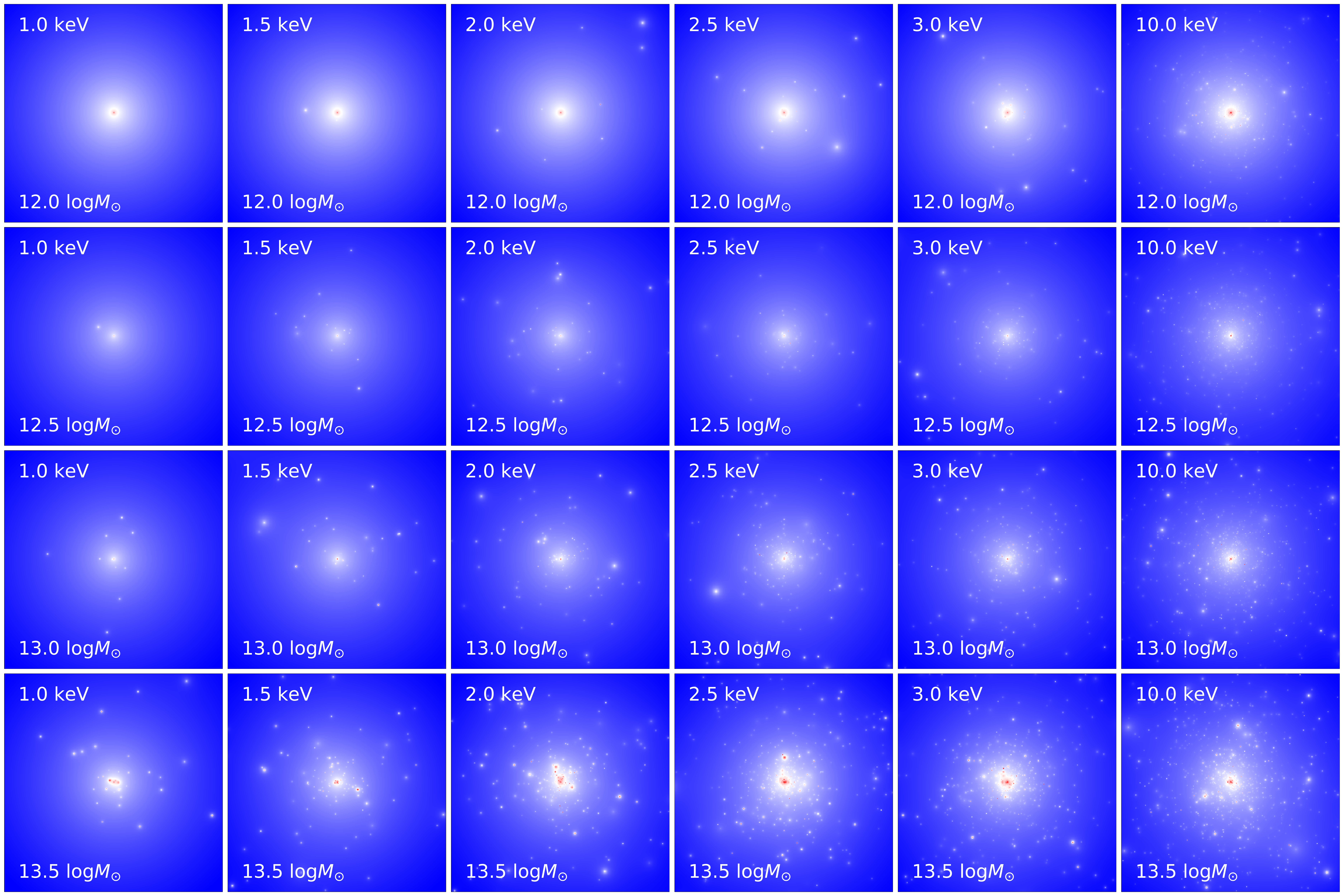}
  \caption{The projected mass of a set of different stochastic semi-analytic substructure realizations. From left to right: Increasing dark matter thermal relic mass from 1 keV to 10 keV. From top to bottom: Increasing parent halo mass from $10^{12}M_{\odot}$-$10^{13.5}M_{\odot}$. The size of the region is identical to the HST image being modeled, i.e. 4" by 4". The color scale is fixed for different dark matter models but change with halo mass.}
  \label{fig:wdm_halos}
\end{figure*}

The prescriptions and assumptions stated here allow us to generate mock halos of different masses with substructure based on different WDM thermal relic masses. Figure \ref{fig:wdm_halos} shows a set of realizations of halos in the range 12.0-13.5 log$(M/M_{\odot})$ and thermal relic masses 1.0-10.0 keV. The enhanced amount of substructure of higher WDM mass is visible on the figure.

\subsection{Mock image generation}
In our approach, we aim to separate the effects of global features from those coming from substructure. In generating mock images, we have therefore decided to use a hybrid approach. For the global features, we use the deflection angles calculated by fitting the smooth lens to the data, $\vec{\alpha}_{\text{smooth}}$. This allows us to generate mock strong lens systems that are in the same regime as that of the data. The small scale features, which is where dark matter properties have the most significant impact, are modeled using the halos generated by the merger tree procedure described in section \ref{sec:simulations}. To do this, we first calculate the full deflection angles for the mock halos $\vec{\alpha}_{dm}$. We then generate a smoothed field $\vec{\alpha}^{s}_{dm}$ through a convolution with a Gaussian kernel $g$ of width $\phi_{\text{min}}$,
\begin{equation}
	\vec{\alpha}^{s}_{dm} = \vec{\alpha}_{dm} \ast g(\phi_{\text{min}}).
\end{equation}

\begin{figure*}
  \centering
  \includegraphics[angle=0, width=145mm]{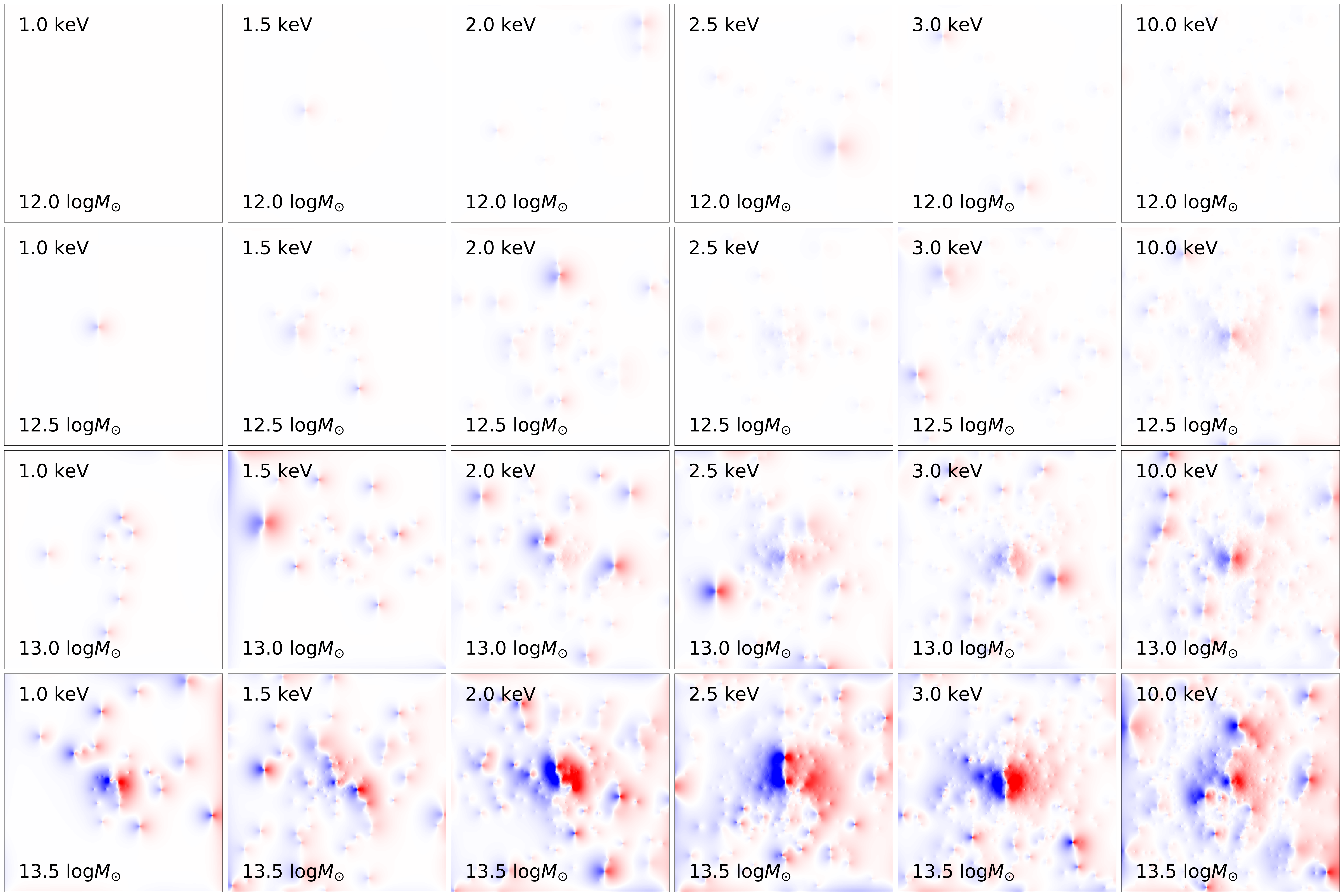}
  \caption{The deflection perturbation $\Delta \alpha_{1, dm}$ of the same set of different stochastic semi-analytic substructure realizations as in figure \ref{fig:wdm_halos}. From left to right: Increasing dark matter thermal relic mass. From top to bottom: Increasing parent halo masses. The the size is again 4" by 4". The color scale is chosen such that full color (red or blue) indicate deflection perturbation of 0.04".}
  \label{fig:wdm_alphas}
\end{figure*}

We then approximate the perturbations to the deflection angles $\Delta \vec{\alpha}_{dm}$ as
\begin{equation}
	\Delta \vec{\alpha}_{dm} = \vec{\alpha}_{dm} - \vec{\alpha}^{s}_{dm}.
\end{equation}
We chose the kernel width $\phi_{\text{min}}$ to match the smooth lens model shapelet deflection scale, $\theta_{\text{min}}= 0.4"$. The result is a deflection angle map that has no net mass beyond scales of $\theta_{\text{min}}$\footnote{There is no unique way to filter scales below $\phi_{\text{min}}$. We also performed a median wavelet transform up to the level matching $\phi_{\text{min}}$ on a subset of the halo realizations. We find no significant impact on the summary statistics we apply. We conclude that the details of the filter does not significantly impact our analysis.}. As an example, Figure \ref{fig:wdm_alphas}  shows the deflection angle perturbation map of one of the components of $\Delta \alpha_{dm}$ for the cases shown in Figure \ref{fig:wdm_halos}.

The deflections used to generate the mock images are therefore
\begin{equation}
	\vec{\alpha}_{\text{mock}} = \vec{\alpha}_{\text{smooth}} + \Delta \vec{\alpha}_{dm} .
\end{equation}

We also use the same source model, PSF model and weight map as inferred from the data and we add Poisson and Gaussian noise on the mock image.

\section{Substructure model comparison} \label{sec:model_comparison}

The tools introduced in the previous sections \ref{sec:simulations} and \ref{sec:smooth_lens_model} allow us to generate mock data that is very similar to the actual imaging data. Our aim is to construct a statistical test that will allow us to select from different dark matter models. The Approximate Bayesian Computing (ABC) \citep[][]{Turner:2012p14944, Liepe:2014p14925, Akeret:2015p15286} framework allows us to construct a posterior distribution even for cases where the likelihood, probability of data given a model, can not be calculated. Instead, ABC relies on detailed forward modeling to capture the impact of errors. It requires distance measures (metric distances) between the output of different simulations, based on summary statistics. In general, the summary statistic can take many forms but needs to map to one or several distance metrics.

Our goal is to construct distance metrics that we can measured from the data and to the mock data and that are sensitive to dark matter properties. This means that comparisons of different realizations of the same dark matter model should result in small distances. On the other hand, comparing realizations of different dark matter models should result in large distances.

The features (substructure in the lens galaxy) that we are targeting emerge due to astrometric anomalies of surface brightness in the reconstructed image compared with a smooth lens model. As a complication, the intrinsic source surface brightness distribution is not known in the actual data. Therefore, anomalies can only be quantified by simultaneously reconstructing the lens (with and without small scale perturbations) and the source.

Instead of trying to identify individual structures within the lens, we use a scanning strategy that introduce local perturbations in the lens model as described below in \ref{sec:substructure_scanning}. These scans provide us with information that we further compress into a summary statistics (\ref{sec:model_comparison_sub}) to discriminate between different models of dark matter with ABC.

\subsection{Substructure scanning procedure}\label{sec:substructure_scanning}
Our scanning procedure works by fitting a lens model that includes a single subclump perturber, with corresponding deflection $\vec{\alpha}_{\text{clump}}$, at a position ($x_i,y_i$) as well as the smooth global model $\vec{\alpha}_{\text{smooth}}$. The combined deflection angle is then
\begin{equation}
	\vec{\alpha}_{\text{pert}}(x_i,y_i) = \vec{\alpha}_{\text{smooth}} + \vec{\alpha}_{\text{clump}}(x_i, y_i).
\end{equation}

We have chosen the perturber to have an SIS profile up to a radius $r_{\text{trunc}}$, followed by a linearly decreasing deflection such that beyond $2r_{\text{trunc}}$, any perturbative deflection disappears. The functional form of the perturber is
\begin{equation}
	\alpha_{\text{clump}}(r) =
	\begin{cases}
    \theta_{\text{E,clump}} & \text{if } r \le r_{\text{trunc}}\\
    \theta_{\text{E,clump}} \left(2 - \frac{r}{r_{\text{trunc}}}\right) & \text{if } r_{\text{trunc}} < r \le 2r_{\text{trunc}}\\
    0 & \text{if } r > 2 r_{\text{trunc}}
	\end{cases}
\end{equation}
where $\theta_{\text{E,clump}}$ is the Einstein radius of the clump and $r_{\text{trunc}}$ its truncation radius. For our analysis, we set $\theta_{\text{E,clump}} = 0.01"$ and $r_{\text{trunc}} = 0.1"$. The total mass enclosed in the truncation radius is $M_{\text{clump}} = 10^{8.35}M_{\odot}$ for the lens in consideration. 

We then move the position of the subclump perturber in increments of 0.05" over the image in both the horizontal and vertical directions. At each position ($x_i,y_i$), we make a best fit reconstruction of the lensed image $I$ (this can be a fit to data or to simulated data) using the perturbed model ($\vec{\alpha}_{\text{pert}}$). We call this reconstructed image $I^{\text{pert}}_i$. We also perform fits using the smooth model $\vec{\alpha}_{\text{smooth}}$ leading to the reconstructed images $I^{\text{smooth}}_i$. When performing a fit we again use the source basis sets as described in section \ref{sec:smooth_lens_model}. In addition, we add source surface brightness shapelets of order $n_{\text{max}} = 5$ at the source position
\begin{equation}
	(x'_i,y'_i) = (x_i,y_i) - \vec{\alpha}_{\text{smooth}}(x_i,y_i),
\end{equation}
which corresponds to the source plane position being mapped through the subclump perturber. We set the scale $\beta$ of these additional shapelets to the average magnification within the truncated SIS profile at ($x_i,y_i$).

\begin{figure*}
  \centering
  \includegraphics[angle=0, width=150mm]{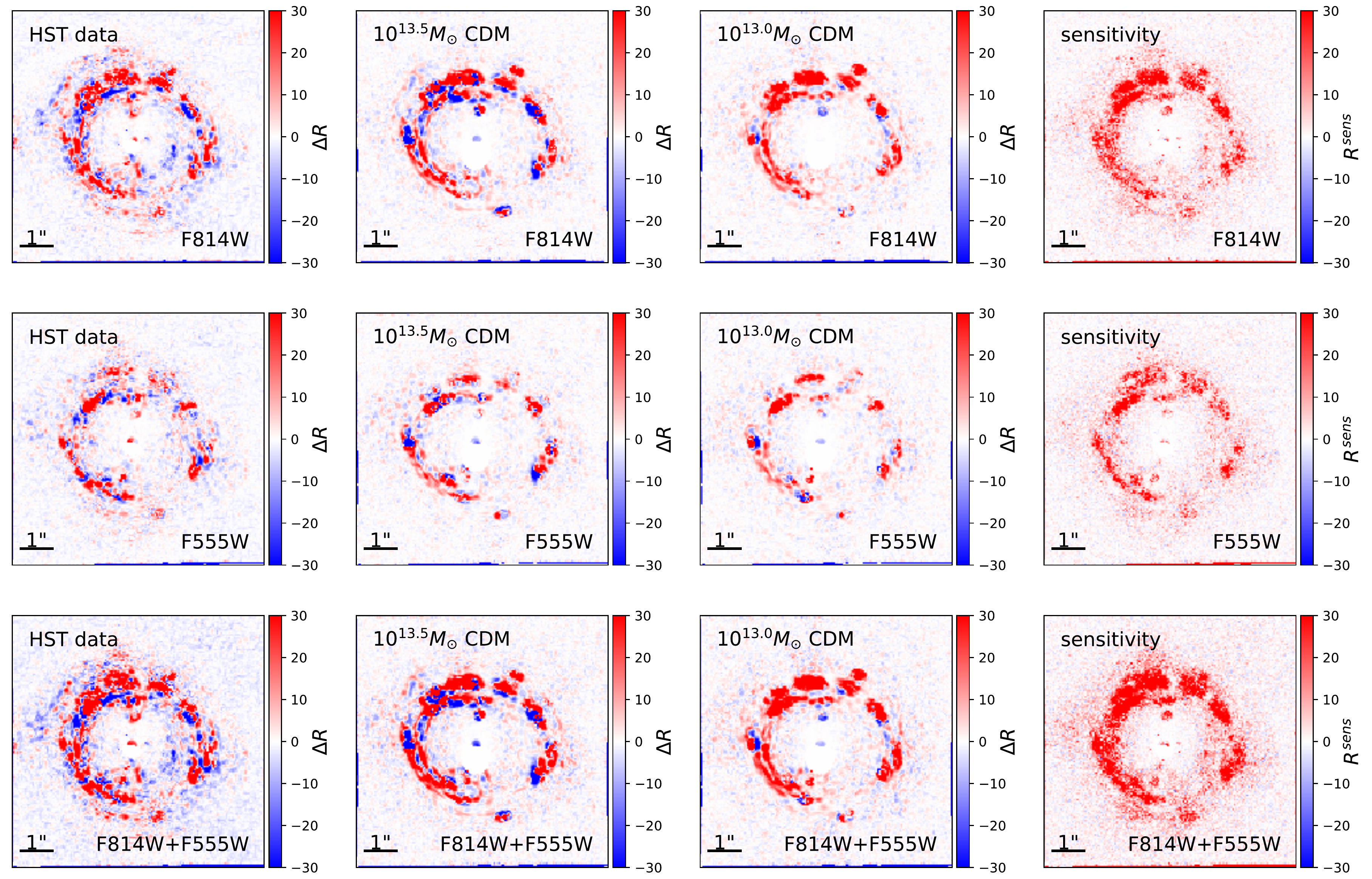}
  \caption{The scanning results for $\Delta R_i$ of the HST data (left column) and two selected CDM semi-analytic realizations with halo masses $10^{13.5}M_{\odot}$ and $10^{13}M_{\odot}$ (middle two columns) and the sensitivity map  (right column). The different rows indicate the analysis of observing band F814W (top), F555W (middle) and combined F814W+F555W (bottom). Each pixel in the plot reflects $\Delta D_i$ when placing the perturber at the position of the pixel.}
  \label{fig:scann_example}
\end{figure*}

From the set of images that we generate, we can define three different residual measures ($R$), related to the relative agreement of the images $I$ (the data or mock generated data), $I^{\text{pert}}_i$ (the best fit reconstruction of $I$ with a smooth model and one additional perturber) and $I^{\text{smooth}}_i$ (the best fit reconstruction of $I$ with a smooth model ), which are effectively a $\chi^2$ measures, as
\begin{equation}
	R^{\text{pert}}_i = \sum_{\text{pixel}} \frac{1}{\sigma^2}\left(I -  I^{\text{pert}}_i\right)^2,
\end{equation}

\begin{equation}
	R^{\text{smooth}}_i = \sum_{\text{pixel}} \frac{1}{\sigma^2}\left(I -  I^{\text{smooth}}_i\right)^2,
\end{equation}
and
\begin{equation}
	R^{\text{sens}}_i = \sum_{\text{pixel}} \frac{1}{\sigma^2}\left(I^{\text{pert}}_i -  I^{\text{smooth}}_i\right)^2,
\end{equation}
where the sum is over the pixels of the image and $\sigma$ is the noise associated with each pixel. We do not expect any of those measures to agree (i.e reduced $\chi^2\approx 1$), i.e neither a smooth model nor a smooth model with one added truncated perturber can describe the data (or simulated data) sufficiently. We expect the sensitivity measure $R^{\text{sens}}_i$ to be the closest to agreement. The two models compared in $R^{\text{sens}}_i$ only differ in the addition of one truncated perturber in one model and the relative residual comes from a very targeted presence of one perturber. This measure provides us with an estimate of the maximal additional residual measure to be excepted in the presence of a subclump. In this regard, to quantify the deflection anomalies at particular positions,
we further introduce a relative excess distance $\Delta R_i$ defined as
\begin{equation} \label{eqn:relative_excess_dist}
	\Delta R_i = R^{\text{pert}}_i - R^{\text{smooth}}_i.
\end{equation}
Negative $\Delta R_i$ show that an addition of a subclump perturber improves the reconstruction of the data (or simulated data) while a positive $\Delta R_i$ shows that the smooth model performs better in reconstructing the original image. Figure \ref{fig:scann_example} shows the scanning results, in terms of $\Delta R$ values, for the HST data (left column), a model with a halo mass of $10^{13.5}M_{\odot}$ (second column), a model with a halo mass of $10^{13.0}M_{\odot}$ (third column). The last column shows the sensitivity map, $R^{\text{sens}}_i$. The different rows indicate the results for different observing bands. Band F814W results are shown in the top row, F555W in the middle and the combined results F814W+F555W at the bottom.

\subsection{Distance metric to compare simulations}\label{sec:model_comparison_sub}

Given the scan maps that we are able to generate for both data and mocks, the challenge is to construct a distance metric that allows us to compare two different scan maps and encapsulate the statistical features imprinted in them. In principle, any metric can be applied to the scann maps to compare two distributions. The challenge is to contract the information as much as possible not to find our self to reject too many simulations while keeping most of the contained information about the quantities of interest. The imprint of dark matter properties is effectively mapped to the abundance and density profile of subhaloes at different masses. These primary quantities of interest in the deflection pattern result in the abundances of deflection anomalies and their spacial patterns in the substructure scans. 

To probe the abundances and masses of the subhalo population imprinted in the deflection anomalies, our primary statistics is the abundance and spacial pattern of negative excess distance $\Delta R_i < 0$ (equation \ref{eqn:relative_excess_dist} described in section \ref{sec:substructure_scanning}). In this work, we restrict ourself to the pattern emerging from $\Delta R_i < -4$, which is more stable to artifacts in the noise modelling of the simulations compared to the data.

We find that feasible distance measures for our aim can be constructed based the spherical averaged two-point correlation function $C(dr) = \left<\Delta R_i(r)\Delta R_i(r+dr)\right>_{r}$ of the negative residuals $\Delta R_i < 0$. This function contains information about the number of anomalies (normalization) and their spatial patterns (slope). We chose the metric to comparing two correlation functions of two realisations of data $C_1(dr)$ and $C_2(dr)$ as
\begin{equation} \label{eqn:corr_dist}
	D(C_1, C_2) = \int_{0"}^{2.5"} \left(C_1(dr) - C_2(dr)\right)^2dr,
\end{equation} 
where the integral goes over scales between 0" and 2.5" with equal weight on all scales.
We emphasise that the expressions in equations \ref{eqn:corr_dist} is not a likelihood. The ABC framework will turn a rejection based on that metric in a likelihood.

\begin{figure*}
  \centering
  \includegraphics[angle=0, width=150mm]{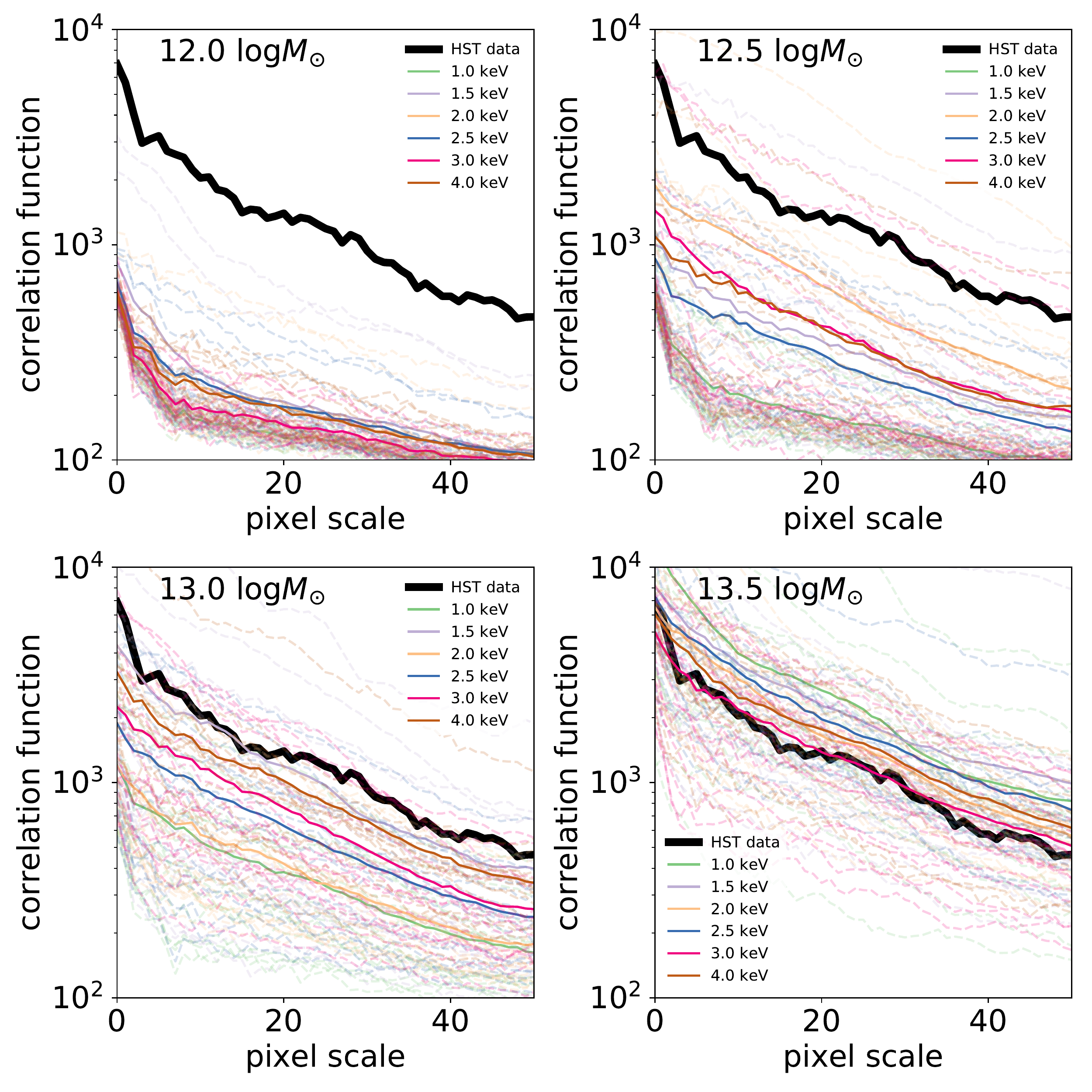}
  \caption{Correlation function of the excess residuals $C(dr)$ of several realisations for six different dark matter models, where the panel are for different parent halo masses. Top left to bottom right: Increasing parent halo mass. Bold black line indicates the distribution of the data. Continuous lines are the mean distribution for the different dark matter models. Dashed lines are individual realizations of the mocks.}
  \label{fig:correlations}
\end{figure*}

Figure \ref{fig:correlations} shows the one-dimensional correlation function $C(dr)$ of several realisations for six different dark matter models, where the panels are for different parent halo masses. The thick black line shows the statistics extracted from the data.

\section{Dark Matter model constraints} \label{sec:dm_constraints}

In this work, we focus on thermal relic mass constraints for dark matter. From the diagnostics of section \ref{sec:model_comparison} and Figure \ref{fig:correlations}, we see that the mass of the parent halo also has a significant impact on our distance metrics and therefore needs to be included as a free parameter in our analysis. Hence, we focus on the two parameters thermal relic mass \mth  and parent halo mass $M_h$. The goal is to calculate the posterior distribution in \mth, marginalized over halo mass $M_h$. We set the prior on the halo mass as flat in log-space in the range [$10^{12}M_{\odot}$, $10^{13.5}M_{\odot}$] and for \mth  uniform in [$1$keV, $5$keV]. For all practical purposes in our simulations and the quality of the data, $\mth > 5$keV corresponds to CDM. To explore the parameter space in an ABC analysis, a large number of simulations are required. 

After an initial grid based study with a total of 480 realisations in the range $M_h \in [10^{12}M_{\odot}, 10^{13.5}M_{\odot}]$ and \mth
$\in [1,10]$keV, we are able to exclude any halo mass below $10^{13}M_{\odot}$ by more than $2\sigma$ confidence\footnote{This statement is sample limited and is confirmed in the more dense sampling of a narrower region in parameter space.} and found that a good prior for our full ABC calculations to be $M_h \in [10^{13}M_{\odot}, 10^{13.5}M_{\odot}]$ and \mth
$\in [1,5]$keV. We then randomly sample from this prior to produce a total of 2483 mock data sets.

\begin{figure*}
  \centering
  \includegraphics[angle=0, width=160mm]{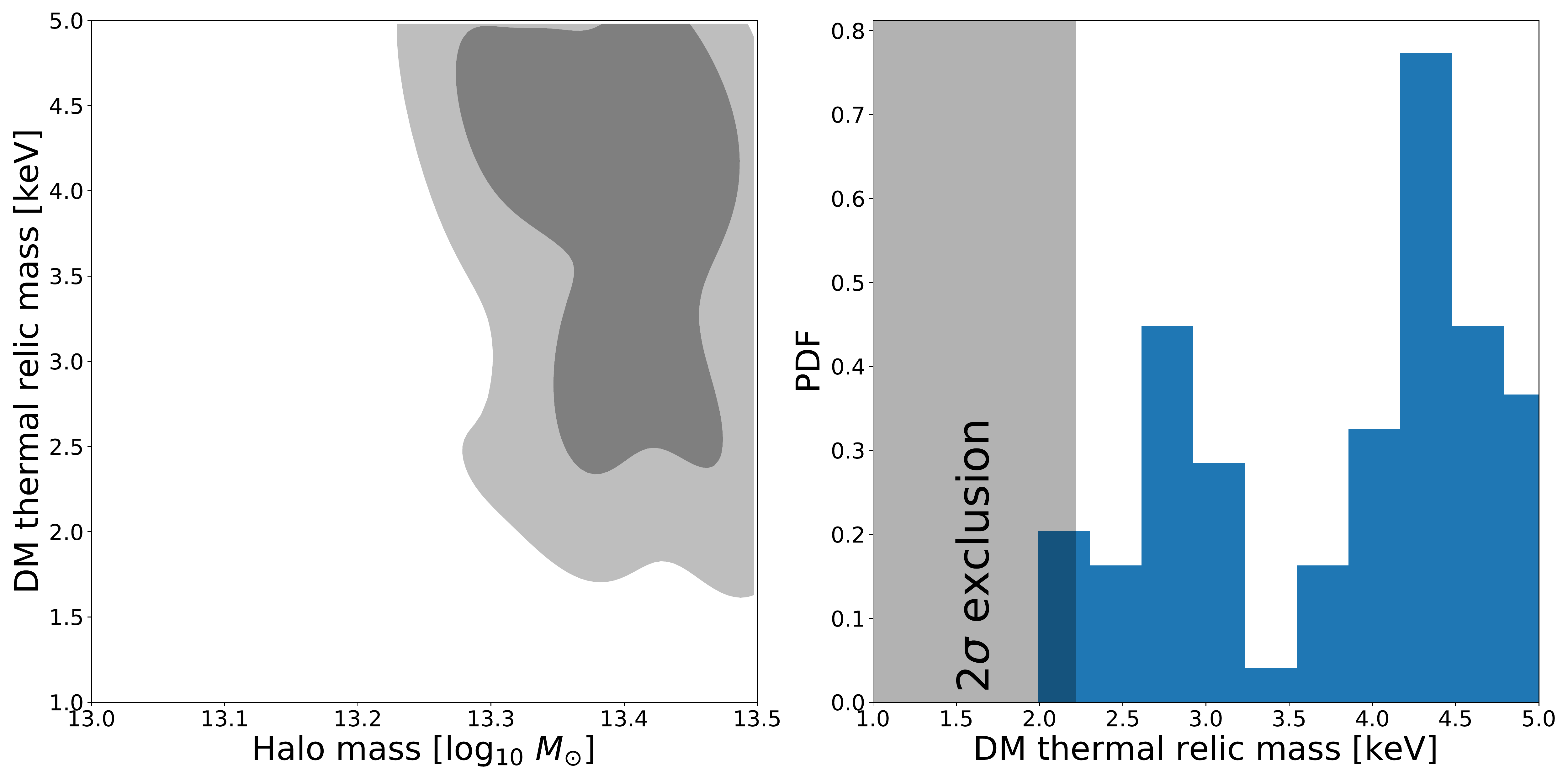}
  \caption{\textbf{Left:} 1-$\sigma$ (dark region) and 2-$\sigma$ (light region) posterior distribution estimated by ABC method on the thermal relic mass vs halo mass plane from the lens \RXJ. \textbf{Right:} The 1-d marginals of the same distribution for the thermal relic mass. The shadowed area indicates the 2-$\sigma$ exclusion region in \mth. The sample number is limited and the details in the posterior distribution is not fully converged.}
  \label{fig:dm_constraints}
\end{figure*}

In the ABC process, we accept the simulations, whose summary statistics in the cumulative distribution and the correlation function reside in the best 5\% of the sample. Setting the threshold in this way does not guarantee that we have reached convergence, however, since lowering the threshold on the distance measures will give tighter constraints if our calculations have not converged, the results we present are conservative estimates. Figure \ref{fig:dm_constraints} (left) shows the two-dimensional distribution of the accepted samples. We see that high halo masses and high thermal relic masses are preferentially selected through our cut in the summary statistics of ABC. Figure \ref{fig:dm_constraints} (right) shows the \mth marginal distribution. We can state that thermal relic mass models below \mkeV keV are disfavored by 2$\sigma$, in agreement with a CDM scenario. This result is comparable in strength and in agreement with the other probes of small scale structure. The latest constraints from the Lyman-$\alpha$ forest \citep[][]{Viel:2013p13930} results in a lower bound on the thermal relic mass $\mth = 3.3$keV at the 2$\sigma$ confidence level. The limits obtained from dwarf galaxy counts dis-favors particle masses below $\mth = 2.3$keV at the 2$\sigma$ confidence level\citep[][]{Polisensky:2011p14007, Kennedy:2014p14040}.

\section{Discussion} \label{sec:discussion_substr}

We have performed a statistical analysis of the substructure content in the strong lens \RXJ and report a lower limit for the mass of thermal relic dark matter of $\mth = \mkeV$keV at 2$\sigma$ confidence level. This is comparable to earlier limits derived from Lyman-$\alpha$ clustering and MW dwarf counts. Our statistical method significantly improves upon clump-by-clump strong lensing analyses and can easily be extended to multiple lens systems. Our forward modelling analysis of \RXJ uses an ABC method and targets statistics that are sensitive measures of dark matter properties.

The model shows that multiple substructures within the same mass range have a significant impact on the lensing statistics. This means that probing substructure on a one-by-one basis at this mass limit may not be feasible. However, when pushing the limits to lower substructure masses, a statistical approach like ours is able to account for the effects of hundreds of subclumps simultaneously. For instance, this can be important for ALMA. ALMA can provide higher resolution data than HST images, which leads to greater potential for measuring smaller lens perturbations \citep[see e.g.][]{Hezaveh:2016p13593, Dye:2015p10118, Rybak:2015p10769} if the information from these higher resolution data can be properly tapped.

Our current constraints are mainly limited by the statistics of the single lens and the moderate sample size of our simulations. The simulations and there analysis requires large computational resources. A larger sample of simulations would allow us to apply stricter cuts in the ABC framework which may lead to tighter constraints. The results we get are therefore conservative estimates of the likelihood. Better data (more lenses, better quality data) can discriminate models with higher significance. Tackling the partial degeneracy between parent halo mass and dark matter thermal relic mass can also be done by incorporating additional, independent, priors on the halo mass. For instance coming from abundance matching or galaxy-galaxy lensing. What we do have from the strong lensing measures is an accurate measure of the total mass (dark matter + baryonic mass) within the Einstein radius, which for \RXJ is $M_{<\theta_E} = 10^{11.9} M_{\odot}$. The host galaxy of the lens is a very massive elliptical early-type galaxy. A significant fraction of the mass within $\theta_E$ comes from baryonic matter. These observations allow us to set a conservative lower limit on the expected halo mass of the lens \RXJ to be $10^{13}M_{\odot}$ from abundance matching and forward modeling of the galaxy population through cosmic time \citep[e.g.][]{Behroozi:2013p7464, Birrer:2014p11555}.

Systematics in the lens modeling could be another limiting factor. In this work, we focused on the source reconstruction scale and the intermediate lens model scale descriptions. Further effects that we do not include could mimic lensing substructure effects, such as micro-lensing by stars of the lensing galaxy, luminous structure of dwarf galaxies, substructure displaced along the line-of-sight or dust extinction. For the current constraints, we do not expect a major impact of such effects on the thermal relic mass. For lower mass halos to infer higher thermal relic mass constraints, substructure along the line-of-sight can contribute significantly to the observed substructure signatures \cite{Frenk:2016p15215, Inoue:2016p15472}. Its first order effect is a renormalization of the subclump mass function, which is effectively a renormalization of the parent halo mass in our description, or a weak lensing effect that we capture in the highly parameterized smooth lens model. Our current constraints are marginalized over the parent halo mass within our prior.

Our method can be extended to probe other statistics of lensing substructure, such as mass-concentration relations and disruption efficiencies and might provide a window to probe suggested solutions of the cusp-core discrepancies. In this work, we have fixed the semi-analytic description of the subhalo evolution based on recent literature. Covariances between the thermal relic constraints and other physics, such as tidal disruption of subhaloes have not been investigated in the current work. Systematics in the semi-analytic description might translate to systematics in the thermal relic mass constraints. We do not expect these systematics to dominate our results at the current level of precision.

Such studies are in reach with the current HST data and with a parallel advance in predictive power of the physical models. With the increasing number of discovered strong lens systems and their follow-up campaigns, significant advances can be expected from strong lens studies in the near future to shed light on the nature of dark matter and the disruption processes of satellite halos.

\section*{Acknowledgments}
SB thanks Aurel Schneider for useful discussions. We acknowledge the import, partial use or inspiration of the following python packages: CosmoHammer \citep[][]{Akeret:2013p8319}, FASTELL \citep[][]{Barkana:1998p5324}, numpy \footnote{www.numpy.org}, scipy \footnote{www.scipy.org}, astropy \footnote{www.astropy.org}, matplotlib \citep[][]{Hunter:2007p14386}, FASTELL \citep[][]{Barkana:1998p5324}, triangle \footnote{https://github.com/dfm/triangle.py}.

\bibliography{papers_bibtex}{}

\providecommand{\href}[2]{#2}\begingroup\raggedright\begin{thebibliography}{10}

\bibitem{PlanckCollaboration:2015p9875}
Planck and P.~A.~R. Ade, \emph{Planck 2015 results. xiii. cosmological
  parameters}, {\emph{arXiv} {\bf astro-ph.CO} (Feb, 2015) },
  [\href{http://arxiv.org/abs/1502.01589v2}{{\tt 1502.01589v2}}].

\bibitem{Anderson:2014p13626}
L.~Anderson, {\'E}.~Aubourg, S.~Bailey, F.~Beutler, V.~Bhardwaj, M.~Blanton
  et~al., \emph{The clustering of galaxies in the sdss-iii baryon oscillation
  spectroscopic survey: baryon acoustic oscillations in the data releases 10
  and 11 galaxy samples},
  \href{http://dx.doi.org/10.1093/mnras/stu523}{\emph{Monthly Notices of the
  Royal Astronomical Society} {\bf 441} (May, 2014) 24}.

\bibitem{Heymans:7p13831}
C.~Heymans, E.~Grocutt, A.~Heavens, M.~Kilbinger, T.~Kitching, F.~Simpson
  et~al., \emph{Cfhtlens tomographic weak lensing cosmological parameter
  constraints: Mitigating the impact of intrinsic galaxy alignments},
  \href{http://dx.doi.org/doi:10.1093/mnras/stt601}{\emph{Monthly Notices of
  the Royal Astronomical Society} {\bf 432} (Dec, 7) 2433--2433--2453--2453}.

\bibitem{TheDarkEnergySurveyCollaboration:2015p13765}
T.~D. E.~S. Collaboration, T.~Abbott, F.~B. Abdalla, S.~Allam, A.~Amara,
  J.~Annis et~al., \emph{Cosmology from cosmic shear with des science
  verification data}, {\emph{eprint arXiv} {\bf 1507} (Jun, 2015)
  arXiv:1507.05552}.

\bibitem{Riess:1998p14437}
A.~G. Riess, A.~V. Filippenko, P.~Challis, A.~Clocchiatti, A.~Diercks, P.~M.
  Garnavich et~al., \emph{Observational evidence from supernovae for an
  accelerating universe and a cosmological constant},
  \href{http://dx.doi.org/10.1086/300499}{\emph{The Astronomical Journal} {\bf
  116} (Aug, 1998) 1009}.

\bibitem{Perlmutter:1999p14481}
S.~Perlmutter, G.~Aldering, G.~Goldhaber, R.~A. Knop, P.~Nugent, P.~G. Castro
  et~al., \emph{Measurements of $\omega$ and $\lambda$ from 42
  high-redshift supernovae}, \href{http://dx.doi.org/10.1086/307221}{\emph{The
  Astrophysical Journal} {\bf 517} (May, 1999) 565}.

\bibitem{Haan:2016p15036}
T.~de~Haan, B.~A. Benson, L.~E. Bleem, S.~W. Allen, D.~E. Applegate, M.~L.~N.
  Ashby et~al., \emph{Cosmological constraints from galaxy clusters in the 2500
  square-degree spt-sz survey}, {\emph{arXiv} {\bf astro-ph.CO} (Mar, 2016) },
  [\href{http://arxiv.org/abs/1603.06522v1}{{\tt 1603.06522v1}}].

\bibitem{Viel:2005p13838}
M.~Viel, J.~Lesgourgues, M.~G. Haehnelt, S.~Matarrese and A.~Riotto,
  \emph{Constraining warm dark matter candidates including sterile neutrinos
  and light gravitinos with wmap and the lyman-α forest},
  \href{http://dx.doi.org/10.1103/PhysRevD.71.063534}{\emph{Physical Review D}
  {\bf 71} (Feb, 2005) 063534}.

\bibitem{Seljak:2006p9712}
U.~Seljak, A.~Slosar and P.~McDonald, \emph{Cosmological parameters from
  combining the lyman-α forest with cmb, galaxy clustering and sn
  constraints},
  \href{http://dx.doi.org/10.1088/1475-7516/2006/10/014}{\emph{Journal of
  Cosmology and Astroparticle Physics} {\bf 10} (Sep, 2006) 014}.

\bibitem{Kauffmann:1993p5842}
G.~Kauffmann, S.~D.~M. White and B.~Guiderdoni, \emph{The formation and
  evolution of galaxies within merging dark matter haloes}, {\emph{Monthly
  Notices of the Royal Astronomical Society} {\bf 264} (Aug, 1993) 201}.

\bibitem{Klypin:1999p4649}
A.~Klypin, A.~V. Kravtsov, O.~Valenzuela and F.~Prada, \emph{Where are the
  missing galactic satellites?},
  \href{http://dx.doi.org/10.1086/307643}{\emph{The Astrophysical Journal} {\bf
  522} (Aug, 1999) 82}.

\bibitem{Moore:1999p4657}
B.~Moore, T.~Quinn, F.~Governato, J.~Stadel and G.~Lake, \emph{Cold collapse
  and the core catastrophe},
  \href{http://dx.doi.org/10.1046/j.1365-8711.1999.03039.x}{\emph{Monthly
  Notices of the Royal Astronomical Society} {\bf 310} (Nov, 1999) 1147}.

\bibitem{Kravtsov:2010p14006}
A.~Kravtsov, \emph{The dark matter annihilation signal from dwarf galaxies and
  subhalos}, \href{http://dx.doi.org/doi:10.1155/2010/281913}{\emph{Advances in
  Astronomy} {\bf 2010} (Dec, 2010) --}.

\bibitem{BoylanKolchin:2011p4244}
M.~Boylan-Kolchin, J.~S. Bullock and M.~Kaplinghat, \emph{Too big to fail? the
  puzzling darkness of massive milky way subhaloes},
  \href{http://dx.doi.org/10.1111/j.1745-3933.2011.01074.x}{\emph{arXiv} {\bf
  astro-ph.CO} (Feb, 2011) }, [\href{http://arxiv.org/abs/1103.0007v2}{{\tt
  1103.0007v2}}].

\bibitem{DelPopolo:2016p14495}
A.~D. Popolo, \emph{On the dark matter haloes inner structure and galaxy
  morphology},
  \href{http://dx.doi.org/10.1007/s10509-016-2810-4}{\emph{Astrophysics and
  Space Science} {\bf 361} (Jun, 2016) 222}.

\bibitem{Sand:2003p15039}
D.~J. Sand, T.~Treu, G.~P. Smith and R.~S. Ellis, \emph{The dark matter
  distribution in the central regions of galaxy clusters},
  \href{http://dx.doi.org/10.1086/382146}{\emph{arXiv} {\bf astro-ph} (Oct,
  2003) }, [\href{http://arxiv.org/abs/astro-ph/0310703v1}{{\tt
  astro-ph/0310703v1}}].

\bibitem{Bode:2001p13951}
P.~Bode, J.~P. Ostriker and N.~Turok, \emph{Halo formation in warm dark matter
  models}, \href{http://dx.doi.org/10.1086/321541}{\emph{The Astrophysical
  Journal} {\bf 556} (Jun, 2001) 93}.

\bibitem{Abazajian:2006p13972}
K.~Abazajian, \emph{Linear cosmological structure limits on warm dark matter},
  \href{http://dx.doi.org/10.1103/PhysRevD.73.063513}{\emph{Physical Review D}
  {\bf 73} (Feb, 2006) 063513}.

\bibitem{Viel:2013p13930}
M.~Viel, G.~D. Becker, J.~S. Bolton and M.~G. Haehnelt, \emph{Warm dark matter
  as a solution to the small scale crisis: New constraints from high redshift
  lyman-α forest data},
  \href{http://dx.doi.org/10.1103/PhysRevD.88.043502}{\emph{Physical Review D}
  {\bf 88} (Jul, 2013) 043502}.

\bibitem{Irsic:2017p15429}
V.~Ir{\v s}i{\v c}, M.~Viel, M.~G. Haehnelt, J.~S. Bolton, S.~Cristiani, G.~D.
  Becker et~al., \emph{New constraints on the free-streaming of warm dark
  matter from intermediate and small scale lyman-$\alpha$ forest data},
  {\emph{eprint arXiv} {\bf 1702} (Feb, 2017) arXiv:1702.01764}.

\bibitem{Polisensky:2011p14007}
E.~Polisensky and M.~Ricotti, \emph{Constraints on the dark matter particle
  mass from the number of milky way satellites},
  \href{http://dx.doi.org/10.1103/PhysRevD.83.043506}{\emph{Physical Review D}
  {\bf 83} (Jan, 2011) 043506}.

\bibitem{Kennedy:2014p14040}
R.~Kennedy, C.~Frenk, S.~Cole and A.~Benson, \emph{Constraining the warm dark
  matter particle mass with milky way satellites},
  \href{http://dx.doi.org/10.1093/mnras/stu719}{\emph{Monthly Notices of the
  Royal Astronomical Society} {\bf 442} (Jul, 2014) 2487}.

\bibitem{Metcalf:2001p9744}
R.~B. Metcalf and P.~Madau, \emph{Compound gravitational lensing as a probe of
  dark matter substructure within galaxy halos},
  \href{http://dx.doi.org/10.1086/323695}{\emph{The Astrophysical Journal} {\bf
  563} (Nov, 2001) 9}.

\bibitem{Keeton:2009p9737}
C.~R. Keeton and L.~A. Moustakas, \emph{A new channel for detecting dark matter
  substructure in galaxies: Gravitational lens time delays},
  \href{http://dx.doi.org/10.1088/0004-637X/699/2/1720}{\emph{The Astrophysical
  Journal} {\bf 699} (Jun, 2009) 1720}.

\bibitem{Moustakas:2009p9694}
L.~A. Moustakas, K.~Abazajian, A.~Benson, A.~S. Bolton, J.~S. Bullock, J.~Chen
  et~al., \emph{Strong gravitational lensing probes of the particle nature of
  dark matter}, {\emph{Astro2010: The Astronomy and Astrophysics Decadal
  Survey} {\bf 2010} (Dec, 2009) 214}.

\bibitem{Koopmans:2005p8841}
L.~V.~E. Koopmans, \emph{Gravitational imaging of cold dark matter
  substructures},
  \href{http://dx.doi.org/10.1111/j.1365-2966.2005.09523.x}{\emph{Monthly
  Notices of the Royal Astronomical Society} {\bf 363} (Oct, 2005) 1136}.

\bibitem{Vegetti:2009p9255}
S.~Vegetti and L.~V.~E. Koopmans, \emph{Bayesian strong gravitational-lens
  modelling on adaptive grids: objective detection of mass substructure in
  galaxies},
  \href{http://dx.doi.org/10.1111/j.1365-2966.2008.14005.x}{\emph{Monthly
  Notices of the Royal Astronomical Society} {\bf 392} (Dec, 2009) 945}.

\bibitem{Vegetti:2010p9515}
S.~Vegetti, L.~V.~E. Koopmans, A.~Bolton, T.~Treu and R.~Gavazzi,
  \emph{Detection of a dark substructure through gravitational imaging},
  \href{http://dx.doi.org/10.1111/j.1365-2966.2010.16865.x}{\emph{Monthly
  Notices of the Royal Astronomical Society} {\bf 408} (Oct, 2010) 1969}.

\bibitem{Vegetti:2012p4937}
S.~Vegetti, D.~J. Lagattuta, J.~P. McKean, M.~W. Auger, C.~D. Fassnacht and
  L.~V.~E. Koopmans, \emph{Gravitational detection of a low-mass dark satellite
  galaxy at cosmological distance},
  \href{http://dx.doi.org/10.1038/nature10669}{\emph{Nature} {\bf 481} (Dec,
  2012) 341}.

\bibitem{Hezaveh:2016p13593}
Y.~D. Hezaveh, N.~Dalal, D.~P. Marrone, Y.-Y. Mao, W.~Morningstar, D.~Wen
  et~al., \emph{Detection of lensing substructure using alma observations of
  the dusty galaxy sdp.81},
  \href{http://dx.doi.org/10.3847/0004-637X/823/1/37}{\emph{The Astrophysical
  Journal} {\bf 823} (Apr, 2016) 37}.

\bibitem{Metcalf:2002p9547}
R.~B. Metcalf and H.~Zhao, \emph{Flux ratios as a probe of dark substructures
  in quadruple-image gravitational lenses},
  \href{http://dx.doi.org/10.1086/339798}{\emph{The Astrophysical Journal} {\bf
  567} (Feb, 2002) L5}.

\bibitem{Amara:2006p4715}
A.~Amara, R.~B. Metcalf, T.~J. Cox and J.~P. Ostriker, \emph{Simulations of
  strong gravitational lensing with substructure},
  \href{http://dx.doi.org/10.1111/j.1365-2966.2006.10053.x}{\emph{Monthly
  Notices of the Royal Astronomical Society} {\bf 367} (Mar, 2006) 1367}.

\bibitem{Metcalf:2012p5467}
R.~B. Metcalf and A.~Amara, \emph{Small-scale structures of dark matter and
  flux anomalies in quasar gravitational lenses},
  \href{http://dx.doi.org/10.1111/j.1365-2966.2011.19982.x}{\emph{Monthly
  Notices of the Royal Astronomical Society} {\bf 419} (Jan, 2012) 3414}.

\bibitem{Xu:2015p9516}
D.~Xu, D.~Sluse, L.~Gao, J.~Wang, C.~Frenk, S.~Mao et~al., \emph{How well can
  cold dark matter substructures account for the observed radio flux-ratio
  anomalies}, \href{http://dx.doi.org/10.1093/mnras/stu2673}{\emph{Monthly
  Notices of the Royal Astronomical Society} {\bf 447} (Feb, 2015) 3189}.

\bibitem{Nierenberg:2017p15476}
A.~M. Nierenberg, T.~Treu, G.~Brammer, A.~H.~G. Peter, C.~D. Fassnacht, C.~R.
  Keeton et~al., \emph{Probing dark matter substructure in the gravitational
  lens he0435-1223 with the wfc3 grism}, {\emph{eprint arXiv} {\bf 1701} (Jan,
  2017) arXiv:1701.05188}.

\bibitem{Inoue:2015p15467}
K.~T. Inoue, R.~Takahashi, T.~Takahashi and T.~Ishiyama, \emph{Constraints on
  warm dark matter from weak lensing in anomalous quadruple lenses},
  \href{http://dx.doi.org/10.1093/mnras/stv194}{\emph{Monthly Notices of the
  Royal Astronomical Society} {\bf 448} (Apr, 2015) 2704}.

\bibitem{Kacprzak:2012p14788}
T.~Kacprzak, J.~Zuntz, B.~Rowe, S.~Bridle, A.~Refregier, A.~Amara et~al.,
  \emph{Measurement and calibration of noise bias in weak lensing galaxy shape
  estimation},
  \href{http://dx.doi.org/10.1111/j.1365-2966.2012.21622.x}{\emph{Monthly
  Notices of the Royal Astronomical Society} {\bf 427} (Nov, 2012) 2711}.

\bibitem{Mesinger:2011p14795}
A.~Mesinger, S.~Furlanetto and R.~Cen, \emph{21cmfast: a fast, seminumerical
  simulation of the high-redshift 21-cm signal},
  \href{http://dx.doi.org/10.1111/j.1365-2966.2010.17731.x}{\emph{Monthly
  Notices of the Royal Astronomical Society} {\bf 411} (Jan, 2011) 955}.

\bibitem{Reinecke:2006p14798}
M.~Reinecke, K.~Dolag, R.~Hell, M.~Bartelmann and T.~A. En{\ss}lin, \emph{A
  simulation pipeline for the planck mission},
  \href{http://dx.doi.org/10.1051/0004-6361:20053413}{\emph{Astronomy and
  Astrophysics} {\bf 445} (Dec, 2006) 373}.

\bibitem{Peterson:2015p14808}
J.~R. Peterson, J.~G. Jernigan, S.~M. Kahn, A.~P. Rasmussen, E.~Peng, Z.~Ahmad
  et~al., \emph{Simulation of astronomical images from optical survey
  telescopes using a comprehensive photon monte carlo approach},
  \href{http://dx.doi.org/10.1088/0067-0049/218/1/14}{\emph{The Astrophysical
  Journal Supplement Series} {\bf 218} (Apr, 2015) 14}.

\bibitem{Juin:2007p14830}
J.~B. Juin, D.~Yvon, A.~R{\'e}fr{\'e}gier and C.~Y{\`e}che, \emph{Cosmology
  with wide-field sz cluster surveys: selection and systematic effects},
  \href{http://dx.doi.org/10.1051/0004-6361:20054680}{\emph{Astronomy and
  Astrophysics} {\bf 465} (Mar, 2007) 57}.

\bibitem{Pires:2009p14840}
S.~Pires, J.-L. Starck, A.~Amara, R.~Teyssier, A.~R{\'e}fr{\'e}gier and
  J.~Fadili, \emph{Fast statistics for weak lensing (fastlens): fast method for
  weak lensing statistics and map making},
  \href{http://dx.doi.org/10.1111/j.1365-2966.2009.14625.x}{\emph{Monthly
  Notices of the Royal Astronomical Society} {\bf 395} (Apr, 2009) 1265}.

\bibitem{Dietrich:2010p14848}
J.~P. Dietrich and J.~Hartlap, \emph{Cosmology with the shear-peak statistics},
  \href{http://dx.doi.org/10.1111/j.1365-2966.2009.15948.x}{\emph{Monthly
  Notices of the Royal Astronomical Society} {\bf 402} (Jan, 2010) 1049}.

\bibitem{Marian:2012p14852}
L.~Marian, R.~E. Smith, S.~Hilbert and P.~Schneider, \emph{Optimized detection
  of shear peaks in weak lensing maps},
  \href{http://dx.doi.org/10.1111/j.1365-2966.2012.20992.x}{\emph{Monthly
  Notices of the Royal Astronomical Society} {\bf 423} (May, 2012) 1711}.

\bibitem{Heymans:2006p14883}
C.~Heymans, L.~V. Waerbeke, D.~Bacon, J.~Berge, G.~Bernstein, E.~Bertin et~al.,
  \emph{The shear testing programme - i. weak lensing analysis of simulated
  ground-based observations},
  \href{http://dx.doi.org/10.1111/j.1365-2966.2006.10198.x}{\emph{Monthly
  Notices of the Royal Astronomical Society} {\bf 368} (Apr, 2006) 1323}.

\bibitem{Massey:2007p14886}
R.~Massey, C.~Heymans, J.~Berg{\'e}, G.~Bernstein, S.~Bridle, D.~Clowe et~al.,
  \emph{The shear testing programme 2: Factors affecting high-precision
  weak-lensing analyses},
  \href{http://dx.doi.org/10.1111/j.1365-2966.2006.11315.x}{\emph{Monthly
  Notices of the Royal Astronomical Society} {\bf 376} (Feb, 2007) 13}.

\bibitem{Bridle:2009p14888}
S.~Bridle, J.~Shawe-Taylor, A.~Amara, D.~Applegate, S.~T. Balan, G.~Bernstein
  et~al., \emph{Handbook for the great08 challenge: An image analysis
  competition for cosmological lensing},
  \href{http://dx.doi.org/10.1214/08-AOAS222}{\emph{Annals of Applied
  Statistics} {\bf 3} (Dec, 2009) 6}.

\bibitem{Refregier:2014p14915}
A.~Refregier and A.~Amara, \emph{A way forward for cosmic shear: Monte-carlo
  control loops},
  \href{http://dx.doi.org/10.1016/j.dark.2014.01.002}{\emph{Physics of the Dark
  Universe} {\bf 3} (Mar, 2014) 1}.

\bibitem{Bruderer:2016p14919}
C.~Bruderer, C.~Chang, A.~Refregier, A.~Amara, J.~Berg{\'e} and L.~Gamper,
  \emph{Calibrated ultra fast image simulations for the dark energy survey},
  \href{http://dx.doi.org/10.3847/0004-637X/817/1/25}{\emph{The Astrophysical
  Journal} {\bf 817} (Dec, 2016) 25}.

\bibitem{Rubin:1984p15259}
D.~Rubin, \emph{Bayesianly justifiable and relevant frequency calculations for
  the applied statistician}, {\emph{The Annals of Statistics} (Jan, 1984) }.

\bibitem{Diggle:1984p15276}
P.~Diggle and R.~Gratton, \emph{Monte carlo methods of inference for implicit
  statistical models}, {\emph{Journal of the Royal Statistical Society. Series
  B ( {\ldots}} (Jan, 1984) }.

\bibitem{Tavare:1997p15227}
S.~Tavar{\'e}, D.~Balding, R.~Griffiths and P.~Donnelly, \emph{Inferring
  coalescence times from dna sequence data}, {\emph{Genetics} (Jan, 1997) }.

\bibitem{Turner:2012p14944}
B.~Turner and T.~V. Zandt, \emph{A tutorial on approximate bayesian
  computation}, {\emph{Journal of Mathematical Psychology} (Dec, 2012) }.

\bibitem{Liepe:2014p14925}
J.~Liepe, P.~Kirk, S.~Filippi, T.~Toni and C.~Barnes{\ldots}, \emph{A framework
  for parameter estimation and model selection from experimental data in
  systems biology using approximate bayesian computation}, {\emph{Nature
  protocols} (Dec, 2014) }.

\bibitem{Akeret:2015p15286}
J.~Akeret, A.~Refregier, A.~Amara, S.~Seehars and C.~Hasner, \emph{Approximate
  bayesian computation for forward modeling in cosmology},
  \href{http://dx.doi.org/10.1088/1475-7516/2015/08/043}{\emph{Journal of
  Cosmology and Astroparticle Physics} {\bf 08} (Aug, 2015) 043}.

\bibitem{Sluse:2003p8680}
D.~Sluse, J.~Surdej, J.-F. Claeskens, D.~Hutsem{\'e}kers, C.~Jean, F.~Courbin
  et~al., \emph{A quadruply imaged quasar with an optical einstein ring
  candidate: 1rxs j113155.4-123155},
  \href{http://dx.doi.org/10.1051/0004-6361:20030904}{\emph{Astronomy and
  Astrophysics} {\bf 406} (Jun, 2003) L43}.

\bibitem{Birrer:2016p13196}
S.~Birrer, A.~Amara and A.~Refregier, \emph{The mass-sheet degeneracy and
  time-delay cosmography: analysis of the strong lens rxj1131-1231},
  \href{http://dx.doi.org/10.1088/1475-7516/2016/08/020}{\emph{Journal of
  Cosmology and Astroparticle Physics} {\bf 08} (Jul, 2016) 020}.

\bibitem{Birrer:2015p11550}
S.~Birrer, A.~Amara and A.~Refregier, \emph{Gravitational lens modeling with
  basis sets}, \href{http://dx.doi.org/10.1088/0004-637X/813/2/102}{\emph{The
  Astrophysical Journal} {\bf 813} (Oct, 2015) 102}.

\bibitem{Refregier:2003p8153}
A.~Refregier, \emph{Shapelets - i. a method for image analysis},
  \href{http://dx.doi.org/10.1046/j.1365-8711.2003.05901.x}{\emph{Monthly
  Notice of the Royal Astronomical Society} {\bf 338} (Dec, 2003) 35}.

\bibitem{Kennedy:2001p8447}
J.~Kennedy, J.~Kennedy and R.~Eberhart, \emph{Swarm intelligence},
  {\emph{books.google.com} (Dec, 2001) }.

\bibitem{Press:1974p410}
W.~H. Press and P.~Schechter, \emph{Formation of galaxies and clusters of
  galaxies by self-similar gravitational condensation},
  \href{http://dx.doi.org/10.1086/152650}{\emph{Astrophysical Journal} {\bf
  187} (Jan, 1974) 425}.

\bibitem{Bond:1991p341}
J.~R. Bond, S.~Cole, G.~Efstathiou and N.~Kaiser, \emph{Excursion set mass
  functions for hierarchical gaussian fluctuations},
  \href{http://dx.doi.org/10.1086/170520}{\emph{Astrophysical Journal} {\bf
  379} (Sep, 1991) 440}.

\bibitem{Lacey:1993p337}
C.~Lacey and S.~Cole, \emph{Merger rates in hierarchical models of galaxy
  formation}, {\emph{Monthly Notices of the Royal Astronomical Society (ISSN
  0035-8711)} {\bf 262} (May, 1993) 627}.

\bibitem{Parkinson:2008p31}
H.~Parkinson, S.~Cole and J.~Helly, \emph{Generating dark matter halo merger
  trees},
  \href{http://dx.doi.org/10.1111/j.1365-2966.2007.12517.x}{\emph{Monthly
  Notices of the Royal Astronomical Society} {\bf 383} (Dec, 2008) 557}.

\bibitem{Neistein:2008p15323}
E.~Neistein and A.~Dekel, \emph{Constructing merger trees that mimic n-body
  simulations},
  \href{http://dx.doi.org/10.1111/j.1365-2966.2007.12570.x}{\emph{Monthly
  Notices of the Royal Astronomical Society} {\bf 383} (Jan, 2008) 615}.

\bibitem{Tweed:2009p15335}
D.~Tweed, J.~Devriendt, J.~Blaizot, S.~Colombi and A.~Slyz, \emph{Building
  merger trees from cosmological n-body simulations. towards improving galaxy
  formation models using subhaloes},
  \href{http://dx.doi.org/10.1051/0004-6361/200911787}{\emph{Astronomy and
  Astrophysics} {\bf 506} (Nov, 2009) 647}.

\bibitem{Benson:2016p15322}
A.~J. Benson, C.~Cannella and S.~Cole, \emph{Achieving convergence in galaxy
  formation models by augmenting n-body merger trees},
  \href{http://dx.doi.org/10.1186/s40668-016-0016-3}{\emph{Computational
  Astrophysics and Cosmology} {\bf 3} (Aug, 2016) 3}.

\bibitem{Benson:2013p13572}
A.~J. Benson, A.~Farahi, S.~Cole, L.~A. Moustakas, A.~Jenkins, M.~Lovell
  et~al., \emph{Dark matter halo merger histories beyond cold dark matter - i.
  methods and application to warm dark matter},
  \href{http://dx.doi.org/10.1093/mnras/sts159}{\emph{Monthly Notices of the
  Royal Astronomical Society} {\bf 428} (Dec, 2013) 1774}.

\bibitem{Barkana:2001p14092}
R.~Barkana, Z.~Haiman and J.~P. Ostriker, \emph{Constraints on warm dark matter
  from cosmological reionization},
  \href{http://dx.doi.org/10.1086/322393}{\emph{The Astrophysical Journal} {\bf
  558} (Aug, 2001) 482}.

\bibitem{Navarro:1997p8389}
J.~F. Navarro, C.~S. Frenk and S.~D.~M. White, \emph{A universal density
  profile from hierarchical clustering}, {\emph{The Astrophysical Journal} {\bf
  490} (Nov, 1997) 493}.

\bibitem{Wechsler:2002p14192}
R.~H. Wechsler, J.~S. Bullock, J.~R. Primack, A.~V. Kravtsov and A.~Dekel,
  \emph{Concentrations of dark halos from their assembly histories},
  \href{http://dx.doi.org/10.1086/338765}{\emph{The Astrophysical Journal} {\bf
  568} (Feb, 2002) 52}.

\bibitem{Ludlow:2013p2973}
A.~D. Ludlow, J.~F. Navarro, M.~Boylan-Kolchin, P.~E. Bett, R.~E. Angulo, M.~Li
  et~al., \emph{The mass profile and accretion history of cold dark matter
  haloes}, \href{http://dx.doi.org/10.1093/mnras/stt526}{\emph{Monthly Notices
  of the Royal Astronomical Society} {\bf 432} (May, 2013) 1103}.

\bibitem{Duffy:2008p14292}
A.~R. Duffy, J.~Schaye, S.~T. Kay and C.~D. Vecchia, \emph{Dark matter halo
  concentrations in the wilkinson microwave anisotropy probe year 5 cosmology},
  \href{http://dx.doi.org/10.1111/j.1745-3933.2008.00537.x}{\emph{Monthly
  Notices of the Royal Astronomical Society: Letters} {\bf 390} (Sep, 2008)
  L64}.

\bibitem{Schneider:2015p14276}
A.~Schneider, \emph{Structure formation with suppressed small-scale
  perturbations}, \href{http://dx.doi.org/10.1093/mnras/stv1169}{\emph{Monthly
  Notices of the Royal Astronomical Society} {\bf 451} (Jul, 2015) 3117}.

\bibitem{Jiang:2016p14171}
F.~Jiang and F.~C. van~den Bosch, \emph{Statistics of dark matter substructure
  - i. model and universal fitting functions},
  \href{http://dx.doi.org/10.1093/mnras/stw439}{\emph{Monthly Notices of the
  Royal Astronomical Society} {\bf 458} (Apr, 2016) 2848}.

\bibitem{vandenBosch:2005p14184}
F.~C. van~den Bosch, G.~Tormen and C.~Giocoli, \emph{The mass function and
  average mass-loss rate of dark matter subhaloes},
  \href{http://dx.doi.org/10.1111/j.1365-2966.2005.08964.x}{\emph{Monthly
  Notices of the Royal Astronomical Society} {\bf 359} (Apr, 2005) 1029}.

\bibitem{Giocoli:2008p312}
C.~Giocoli, G.~Tormen and F.~C. van~den Bosch, \emph{The population of dark
  matter subhaloes: mass functions and average mass-loss rates},
  \href{http://dx.doi.org/10.1111/j.1365-2966.2008.13182.x}{\emph{Monthly
  Notices of the Royal Astronomical Society} {\bf 386} (May, 2008) 2135}.

\bibitem{BoylanKolchin:2008p28}
M.~Boylan-Kolchin, C.-P. Ma and E.~Quataert, \emph{Dynamical friction and
  galaxy merging time-scales},
  \href{http://dx.doi.org/10.1111/j.1365-2966.2007.12530.x}{\emph{Monthly
  Notices of the Royal Astronomical Society} {\bf 383} (Dec, 2008) 93}.

\bibitem{Zentner:2005p290}
A.~R. Zentner, A.~A. Berlind, J.~S. Bullock, A.~V. Kravtsov and R.~H. Wechsler,
  \emph{The physics of galaxy clustering. i. a model for subhalo populations},
  \href{http://dx.doi.org/10.1086/428898}{\emph{The Astrophysical Journal} {\bf
  624} (Apr, 2005) 505}.

\bibitem{Penarrubia:2008p14342}
J.~Pe{\~n}arrubia, J.~F. Navarro and A.~W. McConnachie, \emph{The tidal
  evolution of local group dwarf spheroidals},
  \href{http://dx.doi.org/10.1086/523686}{\emph{The Astrophysical Journal} {\bf
  673} (Dec, 2008) 226--240}.

\bibitem{Penarrubia:2010p14353}
J.~Pe{\~n}arrubia, A.~J. Benson, M.~G. Walker, G.~Gilmore, A.~W. McConnachie
  and L.~Mayer, \emph{The impact of dark matter cusps and cores on the
  satellite galaxy population around spiral galaxies},
  \href{http://dx.doi.org/10.1111/j.1365-2966.2010.16762.x}{\emph{Monthly
  Notices of the Royal Astronomical Society} {\bf 406} (Jul, 2010) 1290}.

\bibitem{Dye:2015p10118}
S.~Dye, C.~Furlanetto, A.~M. Swinbank, C.~Vlahakis, J.~W. Nightingale, L.~Dunne
  et~al., \emph{Revealing the complex nature of the strong gravitationally
  lensed system h-atlas j090311.6+003906 using alma}, {\emph{arXiv} {\bf
  astro-ph.GA} (Mar, 2015) }, [\href{http://arxiv.org/abs/1503.08720v2}{{\tt
  1503.08720v2}}].

\bibitem{Rybak:2015p10769}
M.~Rybak, J.~P. McKean, S.~Vegetti, P.~Andreani and S.~D.~M. White, \emph{Alma
  imaging of sdp.81 - i. a pixelated reconstruction of the far-infrared
  continuum emission},
  \href{http://dx.doi.org/10.1093/mnrasl/slv058}{\emph{Monthly Notices of the
  Royal Astronomical Society: Letters} {\bf 451} (Jun, 2015) L40}.

\bibitem{Behroozi:2013p7464}
P.~S. Behroozi, R.~H. Wechsler and C.~Conroy, \emph{The average star formation
  histories of galaxies in dark matter halos from z = 0-8},
  \href{http://dx.doi.org/10.1088/0004-637X/770/1/57}{\emph{The Astrophysical
  Journal} {\bf 770} (May, 2013) 57}.

\bibitem{Birrer:2014p11555}
S.~Birrer, S.~Lilly, A.~Amara, A.~Paranjape and A.~Refregier, \emph{A simple
  model linking galaxy and dark matter evolution},
  \href{http://dx.doi.org/10.1088/0004-637X/793/1/12}{\emph{The Astrophysical
  Journal} {\bf 793} (Aug, 2014) 12}.

\bibitem{Frenk:2016p15215}
C.~S. Frenk, S.~Cole, Q.~Wang and L.~Gao, \emph{Projection effects in the
  strong lensing study of subhaloes}, {\emph{arXiv} {\bf astro-ph.CO} (Dec,
  2016) }, [\href{http://arxiv.org/abs/1612.06227v1}{{\tt 1612.06227v1}}].

\bibitem{Inoue:2016p15472}
K.~T. Inoue, \emph{On the origin of the flux ratio anomaly in quadruple lens
  systems}, \href{http://dx.doi.org/10.1093/mnras/stw1270}{\emph{Monthly
  Notices of the Royal Astronomical Society} {\bf 461} (Sep, 2016) 164}.

\bibitem{Akeret:2013p8319}
J.~Akeret, S.~Seehars, A.~Amara, A.~Refregier and A.~Csillaghy,
  \emph{Cosmohammer: Cosmological parameter estimation with the mcmc hammer},
  {\emph{Astrophysics Source Code Library} (Feb, 2013) 1303.003}.

\bibitem{Barkana:1998p5324}
R.~Barkana, \emph{Fast calculation of a family of elliptical mass gravitational
  lens models}, \href{http://dx.doi.org/10.1086/305950}{\emph{Astrophysical
  Journal v.502} {\bf 502} (Jul, 1998) 531}.

\bibitem{Hunter:2007p14386}
J.~Hunter, \emph{Matplotlib: A 2d graphics environment}, {\emph{Computing in
  science and engineering} (Dec, 2007) }.

\bibitem{Lovell:2012p15140}
M.~R. Lovell, V.~Eke, C.~S. Frenk, L.~Gao, A.~Jenkins, T.~Theuns et~al.,
  \emph{The haloes of bright satellite galaxies in a warm dark matter
  universe},
  \href{http://dx.doi.org/10.1111/j.1365-2966.2011.20200.x}{\emph{Monthly
  Notices of the Royal Astronomical Society} {\bf 420} (Feb, 2012) 2318}.

\end{thebibliography}\endgroup

\appendix

\section{Systematics in the modeling} \label{app:systematics}
Spurious artifacts in the reconstruction modeling may lead to signatures visible in the summary statistics (see section \ref{sec:model_comparison}) and may therefore be interpreted as caused by substructure. In this appendix, we investigate to what extent the resolution of the smooth lens model and on the source surface brightness reconstruction can impact our statistics.

\subsection{Substructure deflection perturbations} \label{app:sys_smooth_lens}
We use a smooth lens model described in section \ref{sec:smooth_lens_model} with multiple shapelet potentials. This basis set enables us to model the intermediate scales of the lens model. Intermediate smooth mass distribution may give rise to signatures in our summary statistics similar to what small clumps may give rise to as they give rise to astrometric and magnification anomalies as well.

\begin{figure}
  \centering
  \includegraphics[angle=0, width=90mm]{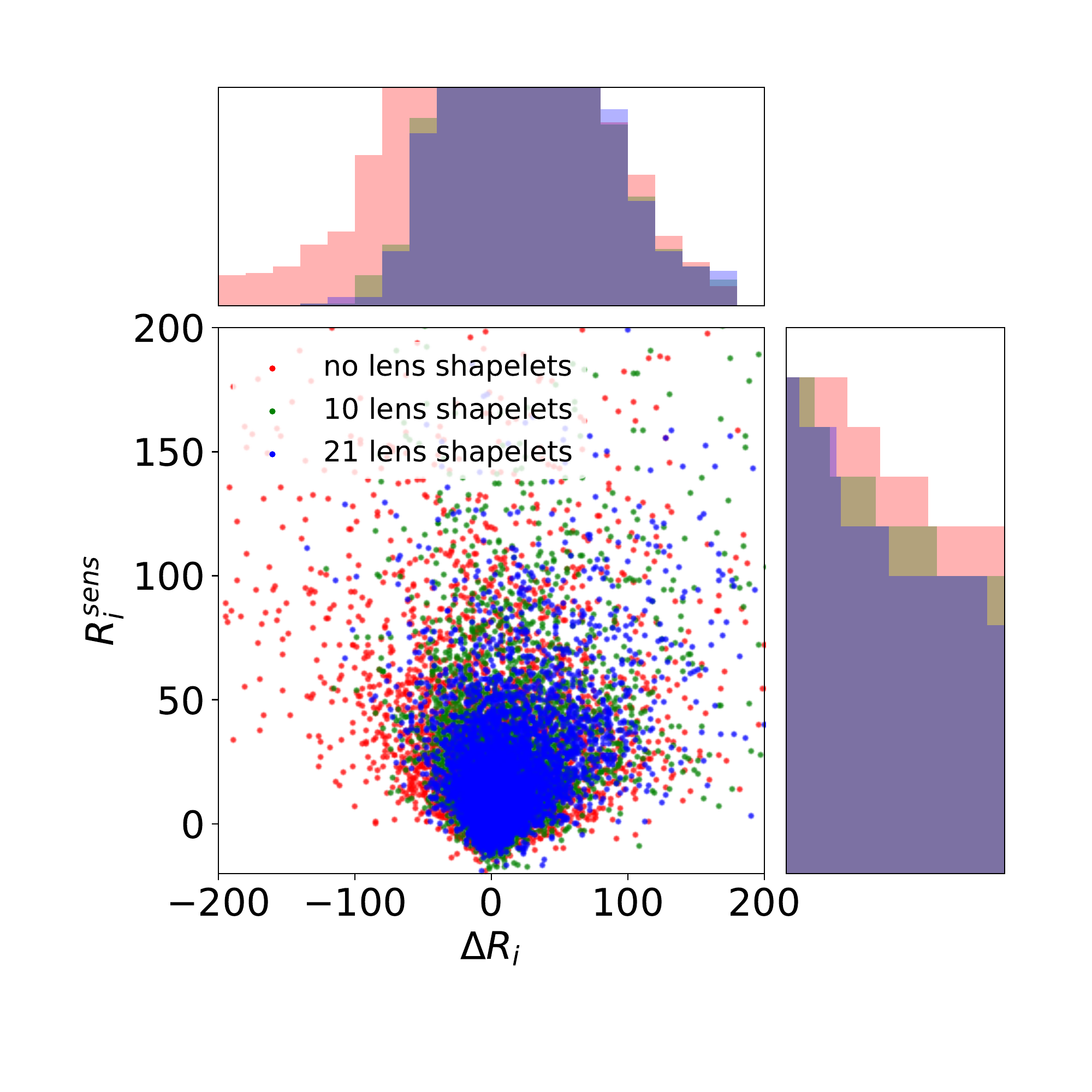}
  \caption{The distributions $\left(\Delta R_i, R^{\text{sens}}_i \right)$ for the data with different resolution in the smooth lens model. Positive y-axis refers to the expected improve in the fit in the presence of the perturber in the mock data. The projected samples are displayed above and on the right.}
  \label{fig:distribution_lens_res}
\end{figure}

To test the impact of intermediate scales on the summary statistics of section \ref{sec:model_comparison}, we model the data with different number of lens model shapelets in addition to the elliptical power-law profile. We fit the data with 0 (no additional intermediate parameterization) 10 ($n_{\text{max}}=3$) and 21 ($n_{\text{max}}=5$) additional shapelets. Figure \ref{fig:distribution_lens_res} shows the $\left(\Delta R_i, R^{\text{sens}}_i \right)$ (section \ref{sec:substructure_scanning}) distributions of those models. We see that the statistics changes significantly when adding the first 10 shapelets. This means that the most of the ``signal" captured in our summary statistics in the smooth model can be attributed to features arising from scales captured by the first 10 lens shapelets. Increasing the shapelet number to 21 does not lead to a significant change in the statistics presented. This means that there are intermediate scales we have to explicitly model before applying our summary statistics. Our underlying assumption is that the features present in the scanning of the model with 21 shapelets arise from scales smaller than being captured by the lens shapelets.

\subsection{Source surface brightness} \label{app:sys_source_res}
The specific source reconstruction model may also have an effect on the summary statistics of section \ref{sec:model_comparison}. Potential signal of substructure relies on the flux and its variation of the source surface brightness. Simulations require to have a source surface brightness very accurately reflecting those of the real data to give rise to the same substructure sensitivities.

\begin{figure}
  \centering
  \includegraphics[angle=0, width=90mm]{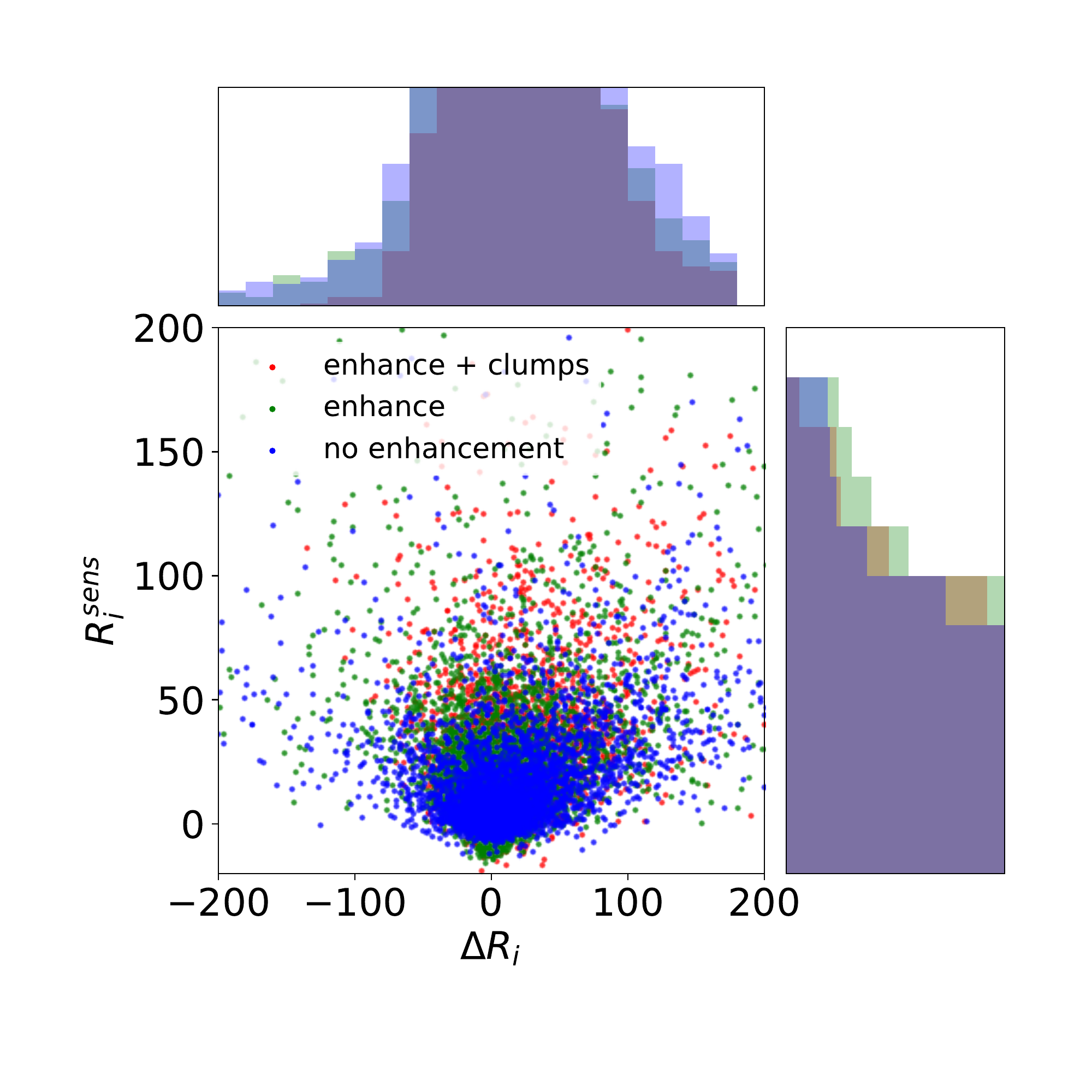}
  \caption{The distributions $\left(\Delta R_i, R^{\text{sens}}_i \right)$ for the data with different resolutions and descriptions of the source surface brightness reconstruction. Negative x-axis refers to improved fit when adding a perturber. Positive y-axis refers to the expected improve in the fit in the presence of the perturber in the mock data. The projected samples are displayed above and on the right.}
  \label{fig:distribution_source_res}
\end{figure}

To investigate this dependence, we take the lens model with 21 additional (our default model) lens shapelets and perform the scanning based on three different source reconstructions. First, we only reconstruct the source with the global shapelets with $n_{\text{max}}=50$. Second, we add the nested shapelet description at the position of the lens perturber (see section \ref{sec:smooth_lens_model}). Third, additionally to the nested shapelets, we add 15 fixed high resolution clumps to the source model (also described in section \ref{sec:smooth_lens_model}). Figure \ref{fig:distribution_source_res} shows the scanning statistics of those three reconstruction models. Substructure perturbation can change the magnification locally significantly. When the source reconstruction can not resolve the existing scales, substructure can significantly help in reconstructing the image because it can demagnify those regions such that the source reconstruction description better match the scales involved. This effect is an artifact. A substructure detection method requires to be able to describe the smallest scales involved in the source surface brightness. We tested our method by further enhancing the nested shapelets and increasing the additionally modeled source clumps. Neither of those pushes to smaller scales did significantly change the substructure scanning statistics. The underlying assumption in our inference is that we are able to match the smallest scales in the source reconstruction description relevant to match the features present in the HST images.

\section{Subhalo statistics}\label{app:subhalo_stat}
\begin{figure}
  \centering
  \includegraphics[angle=0, width=150mm]{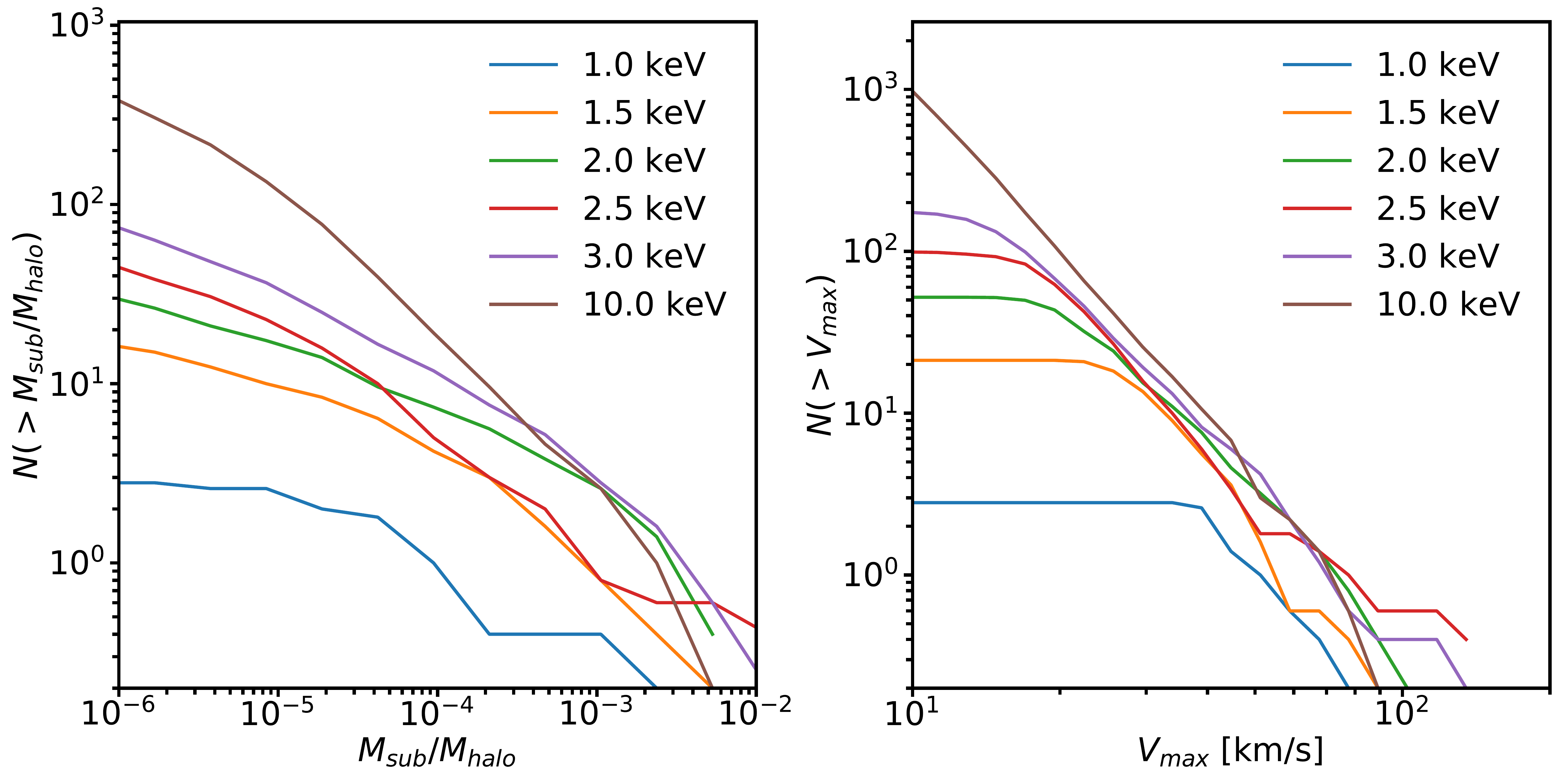}
  \caption{Statistics of the subhalo population generated by our merger tree based formalism for different WDM models. The plots are based on five realizations of a parent halo $M_h = 10^{12.5}M_{\odot}$ at $z=0$. Left: Cumulative subhalo mass function. Right: Cumulative $v_{max}$ distribution of the subhalos.}
  \label{fig:shmf}
\end{figure}

Our semi-analytic description to render a subhalo population described in section \ref{sec:wdm_simulations} provides the statistics of the subhalo population. Figure \ref{fig:shmf} shows the cumulative subhalo mass function and $v_{max}$ function for different WDM models for a parent halo $M_h = 10^{12.5}M_{\odot}$ at $z=0$. The distributions are computed as the average of five realizations of each WDM model. Our results are in good agreement with the N-body simulations of \cite{Lovell:2012p15140}.

\end{document}